\documentclass[english,aps,superscriptaddress,preprintnumbers,nofootinbib,twocolumn]{revtex4}
\usepackage[T1]{fontenc}
\usepackage[latin9]{inputenc}
\usepackage{graphicx}
\usepackage{esint}

\makeatletter
\@ifundefined{textcolor}{}
{%
 \definecolor{BLACK}{gray}{0}
 \definecolor{WHITE}{gray}{1}
 \definecolor{RED}{rgb}{1,0,0}
 \definecolor{GREEN}{rgb}{0,1,0}
 \definecolor{BLUE}{rgb}{0,0,1}
 \definecolor{CYAN}{cmyk}{1,0,0,0}
 \definecolor{MAGENTA}{cmyk}{0,1,0,0}
 \definecolor{YELLOW}{cmyk}{0,0,1,0}
}

\def \tb {\tilde b }

\makeatother

\usepackage{floatrow}
\newfloatcommand{capbtabbox}{table}[][\FBwidth]
\newcommand{\xt}{\mathbf{x}}
\usepackage{babel}

\begin{document}

\title{Off-equilibrium infrared structure of self-interacting scalar fields: Universal scaling, Vortex-antivortex superfluid dynamics \& Bose-Einstein condensation}

\author{Jian Deng}
\affiliation{School of Physics, Shandong University, Jinan, Shandong 250100, China}

\author{Soeren Schlichting}
\affiliation{Department of Physics, University of Washington, Seattle, WA 98195-1560, USA}

\author{Raju Venugopalan}
\affiliation{Brookhaven National Laboratory, Physics Department, Building 510A,
Upton, NY 11973, USA}

\author{Qun Wang}
\affiliation{Interdisciplinary Center for Theoretical Study and Department of
Modern Physics, University of Science and Technology of China, Hefei,
Anhui 230026, China}

\date{\today}
\begin{abstract}

We map the infrared dynamics of a relativistic single component ($N=1$) interacting scalar field theory to that of nonrelativistic complex scalar fields.  The Gross-Pitaevskii (GP) equation, describing the real time dynamics of single component ultracold Bose gases, is obtained at first nontrivial order in an expansion proportional to the powers of $\lambda \phi^2/m^2$ where $\lambda$, $\phi$ and $m$ are the coupling constant, the scalar field and the particle mass respectively. Our analytical studies are corroborated by numerical simulations of the spatial and momentum structure of overoccupied scalar fields in (2+1)-dimensions. 
Universal scaling of infrared modes, vortex-antivortex superfluid dynamics and 
the off-equilibrium formation of a Bose-Einstein condensate are observed. 
Our results for the universal scaling exponents are in agreement with those extracted 
in the numerical simulations of the GP equation. As in these simulations, we observe coarsening phase kinetics in the Bose superfluid with strongly anomalous scaling exponents relative to that of vertex resummed
kinetic theory. Our relativistic field theory  framework further allows one to study more closely the coupling between superfluid and normal fluid modes, specifically the turbulent momentum 
and spatial structure of the coupling between a quasi-particle cascade to 
the infrared and an energy cascade to the ultraviolet. 
We outline possible applications of the formalism to the dynamics of vortex-antivortex formation 
and to the off-equilibrium dynamics of the strongly interacting matter formed in heavy-ion collisions.

\end{abstract}

\preprint{\hfill {\small {ICTS-USTC-18-01}}}
\maketitle

\section{Introdution}

There has been significant progress in recent years in understanding nonequilibrium dynamics in extreme conditions across length scales spanning phenomena as diverse as the early universe shortly after inflation, ultracold quantum gases far from equilibrium and the early time dynamics of ultrarelativistic heavy ion collisions. For a review, see, Ref. \cite{Berges:2015kfa}. Despite the different energy and length scales defining these systems, their temporal evolution can display universal features characteristic of the self-similar dynamics of turbulence.  

Classical-statistical simulations of the self-interacting scalar theories employed to describe both inflationary dynamics as well as ultracold atomic gases far off-equilibrium demonstrate that these systems evolve towards a quasi-stationary nonthermal fixed point (NTFP)~\cite{Berges:2008wm,Berges:2008sr}, where the dynamics is governed by universal scaling behavior~\cite{Micha:2004bv,Orioli:2015dxa}. Besides several investigations in relativistic and nonrelativistic scalar field theories, it has been established that the real time evolution of overoccupied gauge fields (often called the Glasma) also similarly approaches a NTFP~\cite{Berges:2012ev,Schlichting:2012es,Kurkela:2012hp,Berges:2013eia,Berges:2013lsa,Berges:2017igc}. Most strikingly, the nonthermal fixed point realized in non-Abelian gauge theories in an expanding geometry is identical to that seen in overoccupied self-interacting scalars prepared with the same geometry~\cite{Berges:2014bba,Berges:2015ixa}. 

Early investigations of single component ($N=1$) self-interacting scalar field theory performed in the context of inflaton decay~\cite{Khlebnikov:1996mc,Micha:2002ey,Micha:2004bv}, pointed out that the 
emergence of NTFPs can be associated with the transport of conserved quantities (for example the energy) across a large separation of momentum/wave-length scales. 
More recently, it was demonstrate that systems with multiple conserved quantities (such as energy and particle number) 
can realize multiple NTFPs simultaneously in different inertial ranges of momenta~\cite{Berges:2010ez,Berges:2012us,Berges:2013lsa}. One particular example includes $N$-component relativistic scalar theories, where, for strongly overpopulated initial conditions, one finds that in addition to a direct energy cascade towards the ultraviolet (UV)~\cite{Micha:2002ey,Micha:2004bv} there occurs an inverse cascade in quasiparticle number towards the infrared (IR)~\cite{Berges:2012us,Berges:2013lsa,Orioli:2015dxa}. This leads eventually to the formation of an off-equilibrium Bose-Einstein condensate (BEC)~\cite{Berges:2012us,Berges:2013lsa,Orioli:2015dxa}.

Concomitantly, in addition to these numerical investigations, considerable effort has been spent developing an analytic understanding of the universal scaling properties in the vicinity of a NTFP. While it was realized early on that the properties of the UV cascade can be understood in terms of an effective kinetic description~\cite{Micha:2004bv,Berges:2013eia}, this perturbative kinetic theory fails to describe the IR cascade. However, it was argued in \cite{Orioli:2015dxa,Walz:2017ffj} that a vertex-resummed kinetic description based on 
a two particle irreducible (2PI) $1/N$ expansion can simultaneously describe the scaling behavior 
observed in classical statistical simulations in both the IR and UV regimes of relativistic scalar theories~\cite{Orioli:2015dxa}. Further, in this context, it was shown \cite{Orioli:2015dxa} that the statistical properties of the infrared structure of ($N=2$) relativistic scalar fields are identical to that of nonrelativistic scalar fields described by the Gross-Pitaevskii (GP) equation~\cite{Gross:1961,Pitaevskii:1961}.

The GP equation, which is a nonlinear Schr\"{o}dinger equation for the single particle wavefunctions of bosonic atoms, has been studied extensively and exhibits rich structure in both equilibrium~\cite{Pethick:2002,Floerchinger:1900mn,Schmitt:2014eka} and off-equilibrium contexts~\cite{Berloff:2002aa,Svistunov:2015,Nowak:2011sk,Schole:2012kt}. Examples include the equilibrium Berezinskii-Kosterlitz-Thouless (BKT) fixed point characterizing the transition between bound vortex-antivortex pairs and unbound vortices in a two dimensional superfluid~\cite{Berezinskii,Kosterlitz:1973xp}, as well as NTFPs associated with different regimes of weak, strong and superfluid turbulence in the nonequilibrium formation of BECs \cite{Berloff:2002aa,Svistunov:2015,Nowak:2011sk,Schole:2012kt}. Despite the remarkable fact that the vertex-resummed kinetic description of \cite{Orioli:2015dxa,Walz:2017ffj} is able to successfully describe some aspects of the NTPFs in the strong turbulence regime, there is an alternative description of the infrared dynamics of the GP equation in terms of the coarsening dynamics of vortex defects~(see  \cite{Kagan:1994aa,Svistunov:2015}) which is clearly observed in numerical simulations~\cite{Berloff:2002aa,Nowak:2011sk,Schole:2012kt,Karl:2013kua,Karl:2016wko} and in cold atom experiments~\cite{Weiler:2008aa}. Interestingly, there are also examples such as that seen in the evolution of a quenched two-dimensional Bose gas, where the system exhibits ``anomalous'' scaling properties distinct from the 2PI $1/N$ prediction. The former can be understood however from the dynamics of the creation, scattering and decay of vortex-antivortex pairs~\cite{Karl:2016wko}.

The role of topological defects such as strings, domain walls and monopoles, have also been investigated in the context of NTFPs of relativistic scalar theories~\cite{Moore:2015adu,Gasenzer:2011by}. One important insight in this regard concerns the fact that the structure in the deep infrared of $N=1$ relativistic scalars is qualitatively different from those of $N>1$ scalar theories~\cite{Moore:2015adu} as shown by numerical simulations~\cite{Berges:2013lsa,Berges:2017ldx}. Despite the fact that defect structure is different for different number of components $N$, the universal scaling properties of the NTFP appear to be the same for all $N$~\cite{Orioli:2015dxa,Moore:2015adu}. Since there appears to be no natural explanation for this universality in terms of a theory of phase ordering kinetics, it has been unclear thus far to what extent topological defects do or do not play an important role in the description of NTPFs of relativistic scalar field theories.

We will focus our attention on the infrared dynamics of a single component ($N=1$) self-interacting scalar field theory, with the objective of clarifying the relation between relativistic and nonrelativistic scalar field theories and to uncover the role of topological defects in the relativistic case. Although there has been some preliminary discussion of the connection between the infrared dynamics of $N=1$ relativistic scalars and the GP equation~\cite{Berges:2014xea}, a formal identification of the two has been lacking thus far. 
We note that recent attempts to construct an effective field theory (EFT) for the non relativistic dynamics of 
axion dark matter candidates from the underlying relativistic scalar field theory 
\cite{Guth:2014hsa,Braaten:2016kzc,Namjoo:2017nia} bears some similarities to our work. 
In Ref. \cite{Braaten:2016kzc}, an effective field theory for nonrelativistic complex scalar fields 
was constructed by including all terms that satisfy Gallilean invariance, with the coefficients of the EFT determined by matching to the underlying relativistic scalar field theory. 
In Ref. \cite{Namjoo:2017nia}, the nonrelativisitic complex scalar field is, up to 
a factor $e^{imt}$, identical to the field we consider here. 
However Ref. \cite{Namjoo:2017nia} employs a different method from 
our work to to remove non-resonant terms in the effective Lagrangian by exploiting its $U(1)$ symmetry 
corresponding to the conservation of particle number. 
The non-resonant terms carry powers of the phase factor and are 
therefore dropped from the effective action. Ref. \cite{Guth:2014hsa} 
used a simplified definition of the nonrelativistic complex scalar field 
(without non-local operators) as compared with Ref. \cite{Namjoo:2017nia}. 

In this work, we will utilize a Poisson bracket formalism, previously employed in the study of wave turbulence~\cite{Zakharov:1985,Zakharov:1992}, to construct a formal map of the interacting Hamiltonian describing soft modes to a simpler structure in terms of novel canonical variables. This procedure, in the massive relativistic $\lambda \phi^4$ theory that we will study, can be performed order by order in an expansion parameter $\lambda \phi^2/m^2$. In the limit $\lambda \phi^2/m^2 \ll 1$, the GP equation is obtained at the first nontrivial order. The advantage of this procedure is that it allows us in principle to compute higher order corrections to the GP equation in a systematic framework as well as examine within this framework the coupling of superfluid modes to normal fluid modes at higher momenta. 

Our analytical studies are corroborated by off-equilibrium classical-statistical simulations of the $N=1$ weakly coupled relativistic $\lambda \phi^4$ theory. By employing a standard lattice cooling procedure to eliminate hard modes systematically, we will show explicitly that the infrared structure of the relativistic theory exhibits all the characteristic features of the GP equation, including 
the emergence of vortex-antivortex spatial structures and the universal momentum scaling seen in simulations of the GP equation. Our analysis significantly extends earlier numerical~\cite{Orioli:2015dxa} and analytical observations~\cite{Moore:2015adu} which discussed similarities and differences between the two theories. We find in particular that our numerical results for the $N=1$ relativistic scalar theory in (2+1) dimensions are in good agreement with the GP numerical simulations of Karl and Gasenzer~\cite{Karl:2016wko} in the context of a quenched two-dimensional Bose gas. As in Ref. \cite{Karl:2016wko}, we observe different regimes of self-similar infrared behavior that are separated by a characteristic momentum scale corresponding to the typical separation between vortices. We likewise observe the importance of three body collisions of vortex-antivortex pairs with unbound vortices as well as collisions of vortex-antivortex pairs in the coarsening dynamics of the superfluid, leading to ``strongly anomalous''~\cite{Karl:2016wko} scaling properties of the nonthermal fixed point.  These are different from the 2PI $1/N$ predictions of~\cite{Orioli:2015dxa} if extrapolated to the $N=1$ case.


The outline of this paper is as follows. In Section II, we will introduce a Hamiltonian framework for the $N=1$ $\lambda \phi^4$ theory. We then introduce a method of canonical transformations that take advantage of the fact that Poisson brackets are invariant under change of a canonical basis. We show explicitly, at the nontrivial leading order in our expansion parameter, that the Hamiltonian under appropriate canonical transformations will contain only resonant scattering terms. This procedure can be repeated systematically to higher orders, albeit it gets increasingly cumbersome with each increasing order. In Section III, we will show that the canonically transformed quartic Hamiltonian of the $N=1$ $\lambda\phi^4$ theory can be mapped explicitly onto the GP Hamiltonian thereby establishing a direct link between the underlying relativistic scalar field and the GP field theory. In Section IV, we outline the numerical procedure and discuss results from classical statistical simulations of the $\lambda \phi^4$ theory in (2+1) dimensions. In our numerical work, we will  first demonstrate that the map derived analytically in Section III is robust in the infrared, and shall explore its dependence on the mass and the coupling constant.  We next examine the different regimes of self-similar scaling solutions, discuss how one identifies vortex-antivortex structures, compute the vortex density and extract the power law behavior governing its decay. We then study the onset of Bose condensation and relate the temporal power law governing its non-equilibrium onset with the decay of vortex defects. We will conclude with a brief summary and an outline of future related avenues of research. We will in particular speculate on the possible application of these ideas to the study of vorticity in the early time dynamics of heavy-ion collisions. 

The details of computations are given in three appendices. In Appendix \ref{sec:canonical}, we recapitulate general properties of the canonical transformation method discussed in Section \ref{sec:can-var}. In Appendix \ref{sec:coef}, we derive explicitly the coefficients of the auxiliary Hamiltonian that enables the conversion of the $\lambda \phi^4$ Hamiltonian to the resonant Hamiltonian of Section \ref{sec:can-var}. 
In Appendix \ref{sec:dispersion}, for completeness, we discuss some of the properties of the GP equation including the derivation of the superfluid dispersion equation and the superfluid equations of motion. 


\section{Hamiltonian formulation of massive $\lambda \phi^4$ theory}
\label{sec:can-var}
In this section, we will write down the Hamiltonian formulation of the theory in terms of creation-annihilation operators in momentum space. We will then introduce a general method of canonical transformations that will allow us to eliminate nonresonant contributions order by order in powers of the momentum scale of interest over the mass. We will then show in the next section, that the resulting resonant Hamiltonian to quartic order is equivalent to the GP  Hamiltonian. 

We begin with the Lagrangian for $N=1$ massive self-interacting scalar fields: 
\begin{eqnarray}
\mathcal{L} & = & \frac{1}{2}(\partial_{\mu}\phi)(\partial^{\mu}\phi)-\frac{1}{2}m^{2}\phi^{2}-\frac{1}{4!}\lambda\phi^{4}-J\phi.\label{eq:lag-mass}
\end{eqnarray}
We can also understand this Lagrangian as an effective Lagrangian capturing the infrared dynamics of a massless $\lambda {\tilde\phi}^4$, where the field ${\tilde \phi}$ is split into ${\tilde \phi}=\phi+{\tilde \phi}_{\rm hard}$ components. By integrating out the hard modes, one generates for the soft modes $\phi$ an effective mass $m^{2}\equiv (1/2)\lambda\left\langle \tilde{\phi}_{\mathrm{hard}}^{2}\right\rangle$, where $\left\langle \cdots\right\rangle$ denotes an ensemble average over the distribution of initial conditions for the hard modes. One may likewise define $J\equiv (1/6)\lambda\left\langle \tilde{\phi}_{\mathrm{h}}^{3}\right\rangle$. 

Based on the Lagrangian in Eq.~\ref{eq:lag-mass} the Hamiltonian density is given by 
\begin{equation}
\mathcal{H}=\pi\dot{\phi}-\mathcal{L}=\mathcal{H}_{0}+\mathcal{H}_{\mathrm{int}},
\end{equation}
with $\pi\equiv\dot{\phi}$ being the conjugate momentum, and 
\begin{eqnarray}
\mathcal{H}_{0} & = & \frac{1}{2}[\pi^{2}+(\nabla\phi)^{2}+m^{2}\phi^{2}],\nonumber \\
\mathcal{H}_{\mathrm{int}} & = & \frac{1}{24}\lambda\phi^{4}+J\phi. 
\label{h0-h-int}
\end{eqnarray}
Hamiltonian equations of motion are 
\begin{eqnarray}
\frac{d\phi(\mathbf{x})}{dt} & = & \frac{\partial H}{\partial\pi(\mathbf{x})},\nonumber \\
\frac{d\pi(\mathbf{x})}{dt} & = & -\frac{\partial H}{\partial\phi(\mathbf{x})}.\label{eq:ham-eq-1}
\end{eqnarray}
We can express $\phi(\mathbf{x})$ and $\pi(\mathbf{x})$ in terms of their Fourier transforms in momentum space, 
\begin{eqnarray}
\phi(\mathbf{x}) & = & \int[d^{3}\mathbf{k}]\phi_{\mathbf{k}}e^{i\mathbf{k}\cdot\mathbf{x}}\,,\nonumber \\
\pi(\mathbf{x}) & = & \int[d^{3}\mathbf{k}]\pi_{\mathbf{k}}e^{i\mathbf{k}\cdot\mathbf{x}}\,, 
\label{eq:phi-pi}
\end{eqnarray}
where we introduced the shorthand notation $[d^{3}\mathbf{k}]\equiv d^{3}\mathbf{k}/(2\pi)^{3}$. Here and henceforth, all three-vectors are denoted in boldface. 
An on-mass-shell four-momentum is denoted as 
$(E_k,\mathbf{k})$, where the energy is given by $E_k=\sqrt{k^2+m^2}$ 
with the norm of the three-momentum being written as $k\equiv |\mathbf{k}|$. The Fourier transformed $\phi,\pi$ fields can be decomposed as
\begin{eqnarray}
\phi_{\mathbf{k}} & = & \frac{1}{\sqrt{2E_{k}}}(a_{\mathbf{k}}+a_{-\mathbf{k}}^{*})\,,\nonumber \\
\pi_{\mathbf{k}} & = & -i\sqrt{\frac{E_{k}}{2}}(a_{\mathbf{k}}-a_{-\mathbf{k}}^{*})\,,
\label{eq:ak-ask}
\end{eqnarray}
in terms of the complex valued creation-annihilation variables. Expressed in terms of the creation-annihilation variables $a_{\mathbf{k}}$ and $a_{\mathbf{k}}^{*}$, the scalar field Hamiltonian $H=H_{0}+H_{\mathrm{int}}$ takes the form
\begin{eqnarray}
H_{0}(a,a^{*}) & = & \int[d^{3}\mathbf{k}]E_{k}a_{\mathbf{k}}a_{\mathbf{k}}^{*}\,,  \nonumber\\
H_{\mathrm{int}}(a,a^{*})& = & \frac{\lambda}{24}\int \frac{[d^{3}\mathbf{k}][d^{3}\mathbf{k}_{1}][d^{3}\mathbf{k}_{2}][d^{3}\mathbf{k}_{3}]}{\sqrt{16E_{k}E_{k1}E_{k2}E_{k3}}} (2\pi)^{3}\nonumber \\
 &  &\Big[~~~a_{\mathbf{k}}a_{\mathbf{k}1}a_{\mathbf{k}2}a_{\mathbf{k}3}~\delta(\mathbf{k}+\mathbf{k}_{1}+\mathbf{k}_{2}+\mathbf{k}_{3}) \nonumber \\
 & &+4\,a_{\mathbf{k}}a_{\mathbf{k}1}a_{\mathbf{k}2}a_{\mathbf{k}3}^{*}~\delta(\mathbf{k}+\mathbf{k}_{1}+\mathbf{k}_{2}-\mathbf{k}_{3})\nonumber \\
 &  &+6\,a_{\mathbf{k}}a_{\mathbf{k}1}a_{\mathbf{k}2}^{*}a_{\mathbf{k}3}^{*}~\delta(\mathbf{k}+\mathbf{k}_{1}-\mathbf{k}_{2}-\mathbf{k}_{3}) \nonumber \\
 & &+4\,a_{\mathbf{k}}a_{\mathbf{k}1}^{*}a_{\mathbf{k}2}^{*}a_{\mathbf{k}3}^{*}~\delta(\mathbf{k}-\mathbf{k}_{1}-\mathbf{k}_{2}-\mathbf{k}_{3})\nonumber \\
 &  & +~~a_{\mathbf{k}}^{*}a_{\mathbf{k}1}^{*}a_{\mathbf{k}2}^{*}a_{\mathbf{k}3}^{*}~\delta(-\mathbf{k}-\mathbf{k}_{1}-\mathbf{k}_{2}-\mathbf{k}_{3})\Big]\,,\nonumber\\
\label{eq:ham-int}
\end{eqnarray}
where, for convenience, we denote $\delta(\mathbf{k})\equiv\delta^{(3)}(\mathbf{k})$. In the above and in what follows, we have further set the external current $J$ in Eq. (\ref{h0-h-int}) to zero for simplicity. 
Similarly, the equations of motion in Eq.~(\ref{eq:ham-eq-1}) can be rewritten in terms of $a_{\mathbf{p}}$ and $a_{\mathbf{p}}^{*}$ as 
\begin{eqnarray}
\frac{da_{\mathbf{p}}}{dt} & = & -i\frac{\partial H}{\partial a_{\mathbf{p}}^{*}}=-iE_{p}a_{\mathbf{p}}-i\frac{\partial H_{\mathrm{int}}}{\partial a_{\mathbf{p}}^{*}},\nonumber \\
\frac{da_{\mathbf{p}}^{*}}{dt} & = & i\frac{\partial H}{\partial a_{\mathbf{p}}}=iE_{p}a_{\mathbf{p}}^{*}+i\frac{\partial H_{\mathrm{int}}}{\partial a_{\mathbf{p}}}\,.
\label{eq:eom-ap-1}
\end{eqnarray}
This form of the Hamiltonian and the equations of motion will be our starting point for further developments.

We will now adapt the method of canonical variables and canonical transformations, widely used in the study of wave turbulence~\cite{Zakharov:1985,Zakharov:1992}, to our problem. 
The r.h.s of Eq. (\ref{eq:eom-ap-1}) can be written in the form of Poisson brackets as
\begin{eqnarray}
i\frac{da_{\mathbf{p}}}{dt} & = & \left\{ a_{\mathbf{p}},H\right\} _{a}\,,
\end{eqnarray}
where $H$ is given by Eq. (\ref{eq:ham-int}). The subscript $a$ denotes the Poisson bracket in the basis of the canonical variables $a_{\mathbf{p}}$ and $a_{\mathbf{p}}^{*}$  defined by the expression 
\begin{eqnarray}
&& \left\{ F(\mathbf{p}),G(\mathbf{p}_{1})\right\} _{a}  =   \nonumber\\
&& \qquad \int[d^{3}\mathbf{k}]\left[\frac{\partial F(\mathbf{p})}{\partial a_{\mathbf{k}}}\frac{\partial G(\mathbf{p}_{1})}{\partial a_{\mathbf{k}}^{*}}-\frac{\partial F(\mathbf{p})}{\partial a_{\mathbf{k}}^{*}}\frac{\partial G(\mathbf{p}_{1})}{\partial a_{\mathbf{k}}}\right]\,, \nonumber\\
\label{eq:poisson-a}
\end{eqnarray}
where the derivative is taken in the functional sense, namely, $\partial a_{\mathbf{k}}/\partial a_{\mathbf{p}}=(2\pi)^{3}\delta(\mathbf{k}-\mathbf{p})$. One can thereby confirm that the variables $a$ and $a^{*}$ satisfy  the canonical Poisson bracket relations
\begin{eqnarray}
\{a_{\mathbf{p}},a_{\mathbf{p}1}^{*}\}_{a} & = & (2\pi)^{3}\delta(\mathbf{p}-\mathbf{p}_{1}),\nonumber \\
\{a_{\mathbf{p}},a_{\mathbf{p}1}\}_{a} & = & \{a_{\mathbf{p}}^{*},a_{\mathbf{p}1}^{*}\}_{a}=0\,.
\label{eq:poisson-brackets-a}
\end{eqnarray}

Instead of the canonical variables $(a_{\mathbf{k}},a_{\mathbf{k}}^{*})$ we can equivalently use any other set of canonical variables $(b_{\mathbf{p}},b_{\mathbf{p}}^{*})$ to describe the dynamics of the system. Even though the variables $(b_{\mathbf{p}},b_{\mathbf{p}}^{*})$  can be complicated functions of $a_{\mathbf{k}}$ and $a_{\mathbf{k}}^{*}$, the fact that they have to be related by a canonical transformation ensures by definition that  $(b_{\mathbf{p}},b_{\mathbf{p}}^{*})$ satisfy the same Poisson brackets as those 
in the $a$-basis, 
\begin{eqnarray}
\{b_{\mathbf{p}},b_{\mathbf{p}1}^{*}\}_{a} & = & (2\pi)^{3}\delta(\mathbf{p}-\mathbf{p}_{1}),\nonumber \\
\{b_{\mathbf{p}},b_{\mathbf{p}1}\}_{a} & = & \{b_{\mathbf{p}}^{*},b_{\mathbf{p}1}^{*}\}_{a}=0\,.
\label{eq:new-basis}
\end{eqnarray}
One can further prove that the Poisson brackets in Eq.~\ref{eq:poisson-a} are invariant under the change of canonical basis from $(a_{\mathbf{k}},a_{\mathbf{k}}^{*})\rightarrow (b_{\mathbf{p}},b_{\mathbf{p}}^{*})$. This proof is worked out in Appendix \ref{sec:canonical}. One immediate consequence is that the equation of motion for $b_{\mathbf{p}}$ does not change when it is written in the new canonical basis\footnote{In order to avoid confusion we denote the Hamiltonian in terms of the canonical variables $(b_{\mathbf{p}},b_{\mathbf{p}}^{*})$ as $H'$. We note that for time independent canonical transformations $H'(b,b^{*})=H\big(a(b,b^{*}),a^{*}(b,b^{*}) \big)$.}:
\begin{equation}
i\frac{\partial b_{\mathbf{p}}}{\partial t}=\left\{ b_{\mathbf{p}},H'\right\} _{a}=\left\{ b_{\mathbf{p}},H'\right\} _{b}.\label{eq:eom-b-1}
\end{equation}

One particularly elegant way to construct a set of canonical transformations is to generate them via time evolution with an auxiliary Hamiltonian $H_{\mathrm{aux}}$. We emphasize that this \textit{auxiliary time evolution} in the evolution variable $z$ merely corresponds to a change of canonical variables which should not be confused with the physical time evolution of the system. In particular, the Hamiltonian $H_{\mathrm{aux}}$ is {\it a priori} unrelated to the physical Hamiltonian $H$ in Eq.~(\ref{eq:ham-int}). Based on this strategy, we assume that $b_{\mathbf{p}}\equiv {\tilde b}_{\mathbf{p}}(0)$ and $b_{\mathbf{p}}^{*}\equiv {\tilde b}_{\mathbf{p}}^{*}(0)$ are canonical variables at auxiliary time $z=0$, and that $H_{\mathrm{aux}}$ is a real function of ${\tilde b}_{\mathbf{p}}$ and ${\tilde b}_{\mathbf{p}}^{*}$. We can then obtain a new set of canonical variables ${\tilde b}_{\mathbf{p}}(z)$ at any $z>0$ by solving for the auxiliary time evolution: 
\begin{equation}
\label{dbtdz}
i\frac{\partial \tilde{b}_{\mathbf{p}}(z)}{\partial z}=\left\{ \tilde{b}_{\mathbf{p}}(z),H_{\mathrm{aux}}\right\} _{b}\,.
\end{equation}
In practice, the auxiliary time evolution is most conveniently expressed in terms of a Taylor expansion around $z=0$, 
\begin{equation}
{\tb}_{\mathbf{p}}(z) = \sum_{n=0}^{\infty} \frac{z^{n}}{n!} \left.\frac{d^{n} {\tb}_{\mathbf{p}}(z)}{d^{n}z} \right|_{z=0}\,, \nonumber \\
\end{equation}
where the derivatives $d^{n}{\tilde b}_{\mathbf{p}}(z)/d^{n}z$ are given by nested Poisson brackets of ${\tilde b}_{\mathbf{p}}$, ${\tilde b}_{\mathbf{p}}^{*}$
and $H_{\mathrm{aux}}({\tilde b}_{\mathbf{p}},{\tilde b}_{\mathbf{p}}^{*})$ such that
\begin{eqnarray}
 {\tb}_{\mathbf{p}}(z) & = & b_{\mathbf{p}}-iz\{b_{\mathbf{p}},H_{\mathrm{aux}}\}_{b}+\frac{(-iz)^{2}}{2}\big\{\{b_{\mathbf{p}},H_{\mathrm{aux}}\}_{b},H_{\mathrm{aux}}\big\}_{b}\nonumber \\
 &  & +\frac{(-iz)^{3}}{6}\Big\{\big\{\{b_{\mathbf{p}},H_{\mathrm{aux}}\}_{b},H_{\mathrm{aux}}\big\}_{b},H_{\mathrm{aux}}\Big\}_{b}+\cdots, \nonumber \\
\label{eq:taylor-at0}
\end{eqnarray}
and $\tb_{\mathbf{p}}^{*}(z)$ can be obtained by taking the complex conjugate of $\tb_{\mathbf{p}}(z)$. We note that in Eq. (\ref{eq:taylor-at0}), we have set $z=0$ after expressing derivatives in terms of Poisson brackets, such that all $\tb_{\mathbf{p}}$ ($\tb ^*_{\mathbf{p}}$) including those in 
$H_{\mathrm{aux}}({\tilde b}_{\mathbf{p}},{\tilde b}_{\mathbf{p}}^{*})$
become $b_{\mathbf{p}}$ ($b^*_{\mathbf{p}}$) , namely, $\tb_{\mathbf{p}}\rightarrow b_{\mathbf{p}}$, 
$\tb ^*_{\mathbf{p}}\rightarrow b^*_{\mathbf{p}}$, and 
$H_{\mathrm{aux}}({\tilde b}_{\mathbf{p}},{\tilde b}_{\mathbf{p}}^{*})\rightarrow 
H_{\mathrm{aux}}(b_{\mathbf{p}},b_{\mathbf{p}}^{*})$. 
In Appendix \ref{sec:canonical}, we demonstrate explicitly that $\tb_{\mathbf{p}}(z)$
and $\tb_{\mathbf{p}}^{*}(z)$ satisfy the canonical relations in Eq.~(\ref{eq:new-basis}).

%
We will now employ the canonical transformation generated by the auxiliary Hamiltonian $H_{\mathrm{aux}}$ to map our physical Hamiltonian $H(a,a^{*})$ in Eq.~(\ref{eq:ham-int}) expressed in terms of the original variables  $a_{\mathbf{p}}$ and $a_{\mathbf{p}}^{*}$ to the physical Hamiltonian $H^\prime$ expressed in terms of the new canonical variables $b_{\mathbf{p}}$ and $b_{\mathbf{p}}^{*}$. By appropriate choice of $H_{\mathrm{aux}}$ the new Hamiltonian $H'(b,b^{*})$ will have a simpler structure that eliminates so called nonresonant terms with an unequal number of the variables $b_{\mathbf{p}}$ and $b_{\mathbf{p}}^{*}$ corresponding to particle number changing processes. 

We begin by redefining our original variables $a_{\mathbf{p}}$ as ${\tilde b}_{\mathbf{p}}(z)$ (i.e.~$a_{\mathbf{p}}\equiv {\tilde b}_{\mathbf{p}}(z)$) at auxiliary time $z$ and the desired new variable $b_{\mathbf{p}}$ as ${\tilde b}_{\mathbf{p}}(0)$ (i.e.~$b_{\mathbf{p}}\equiv {\tilde b}_{\mathbf{p}}(0)$) at auxiliary time $z=0$. By use of Eq.~(\ref{eq:taylor-at0}), we can then express $a_{\mathbf{p}}$ in terms of a Taylor series as 
\begin{eqnarray}
a_{\mathbf{p}} & = & b_{\mathbf{p}}-iz\{b_{\mathbf{p}},H_{\mathrm{aux}}\}_{b} \nonumber\\
&&+\frac{(-iz)^{2}}{2}\big\{\{b_{\mathbf{p}},H_{\mathrm{aux}}\}_{b},H_{\mathrm{aux}}\big\}_{b}+\cdots\,. \nonumber\\
\label{eq:a-b-z}
\end{eqnarray}
We can obtain $a^*_{\mathbf{p}}$ by taking the complex conjugate of Eq. (\ref{eq:a-b-z}). We choose the auxiliary Hamiltonian $H_{\mathrm{aux}}(b,b^*)$ to be of the form
\begin{eqnarray}
&& H_{\mathrm{aux}}  =   \frac{1}{24} \int[d^{3}\mathbf{k}_{1}][d^{3}\mathbf{k}_{2}][d^{3}\mathbf{k}_{3}][d^{3}\mathbf{k}_{4}]\nonumber \\
 &  &  \left. \times\Big\{B_{1}(\mathbf{k}_{1},\mathbf{k}_{2},\mathbf{k}_{3},\mathbf{k}_{4})b_{\mathbf{k}1}b_{\mathbf{k}2}b_{\mathbf{k}3}b_{\mathbf{k}4}\delta(\mathbf{k}_{1}+\mathbf{k}_{2}+\mathbf{k}_{3}+\mathbf{k}_{4})\right.\nonumber \\
&  & +4 B_{2}(\mathbf{k}_{1},\mathbf{k}_{2},\mathbf{k}_{3};\mathbf{k}_{4})b_{\mathbf{k}1}b_{\mathbf{k}2}b_{\mathbf{k}3}b_{\mathbf{k}4}^{*}\delta(\mathbf{k}_{1}+\mathbf{k}_{2}+\mathbf{k}_{3}-\mathbf{k}_{4})\nonumber \\
&  & +4B_{2}^{*}(\mathbf{k}_{1},\mathbf{k}_{2},\mathbf{k}_{3};\mathbf{k}_{4})b_{\mathbf{k}1}^{*}b_{\mathbf{k}2}^{*}b_{\mathbf{k}3}^{*}b_{\mathbf{k}4}\delta(\mathbf{k}_{1}+\mathbf{k}_{2}+\mathbf{k}_{3}-\mathbf{k}_{4})\nonumber \\
& & + \left. B_{1}^{*}(\mathbf{k}_{1},\mathbf{k}_{2},\mathbf{k}_{3},\mathbf{k}_{4})b_{\mathbf{k}1}^{*}b_{\mathbf{k}2}^{*}b_{\mathbf{k}3}^{*}b_{\mathbf{k}4}^{*}\delta(\mathbf{k}_{1}+\mathbf{k}_{2}+\mathbf{k}_{3}+\mathbf{k}_{4}) \Big\} \right. \,,  \nonumber \\
\label{eq:h-aux}
\end{eqnarray}
where the coefficient functions $B_{1,2}$ have the following properties: (a) $B_{1}(\mathbf{k}_{1},\mathbf{k}_{2},\mathbf{k}_{3},\mathbf{k}_{4})$ is symmetric under any permutation of four momenta; (b) $B_{2}(\mathbf{k}_{1},\mathbf{k}_{2},\mathbf{k}_{3};\mathbf{k}_{4})$ is symmetric for any permutation of first three momenta. 

Now substituting $a_{\mathbf{p}}=\tilde{b}_{\mathbf{p}}(z)$ 
and $a^*_{\mathbf{p}}=\tilde{b}^*_{\mathbf{p}}(z)$ from Eq. (\ref{eq:a-b-z})
into Eq. (\ref{eq:ham-int}) we obtain the Hamiltonian in the new variables $b_{\mathbf{p}}$
and $b_{\mathbf{p}}^{*}$ as 
\begin{eqnarray}
H^{\prime} & = & H(b,b^{*})+(-iz)\{H(b,b^{*}),H_{\mathrm{aux}}\}_{b} \nonumber \\
 & & +\frac{(-iz)^{2}}{2}\big\{\{H(b,b^{*}),H_{\mathrm{aux}}\}_{b},H_{\mathrm{aux}}\big\}_{b}\nonumber \\
 &  & +\frac{(-iz)^{3}}{6}\Big\{\big\{\{H(b,b^{*}),H_{\mathrm{aux}}\}_{b},H_{\mathrm{aux}}\big\}_{b},H_{\mathrm{aux}}\Big\}_{b}\nonumber \\
 & & +\cdots, 
\label{eq:h-new}
\end{eqnarray}
where $H(b,b^{*})=H_0(b,b^{*})+H_{\mathrm{int}}(b,b^{*})$ can be obtained from $H$ in Eq. (\ref{eq:ham-int}) by making the replacement $a_{\mathbf{p}}\rightarrow b_{\mathbf{p}}$ and $a_{\mathbf{p}}^{*}\rightarrow b_{\mathbf{p}}^{*}$. We observe that in the new variables $(b_{\mathbf{p}},b_{\mathbf{p}}^{*})$, the Hamiltonian $H^{\prime}$ behaves as a (auxiliary) time-shift of the Hamiltonian $H(b,b^*)$
whose time evolution is governed by $H_{\mathrm{aux}}$. The proof of Eq. (\ref{eq:h-new}) is given in Appendix \ref{sec:canonical}. Expanding $H^{\prime}$ by using the explicit form of $H(b,b^{*})=H_{0}(b,b^{*})+H_{\mathrm{int}}(b,b^{*})$, we obtain 
\begin{eqnarray}
H^{\prime} & = & H_{0}+H_{\mathrm{int}}+(-iz)\{H_{0}+H_{\mathrm{int}},\; H_{\mathrm{aux}}\}_{b}\nonumber \\
 &  & +\frac{(-iz)^{2}}{2}\big\{\{H_{0}+H_{\mathrm{int}},\; H_{\mathrm{aux}}\}_{b},\; H_{\mathrm{aux}}\big\}_{b}+\cdots\nonumber \\
 \end{eqnarray}
 such that upon grouping the terms of $H^{\prime}$ according to powers of $b_{\mathbf{p}}$ or $b_{\mathbf{p}}^{*}$ involved we obtain
 \begin{eqnarray}
 \label{H-prime}
& &H^{\prime}   =  H_{0}+\left[H_{\mathrm{int}}+(-iz)\{H_{0},H_{\mathrm{aux}}\}_{b}\right]\nonumber \\
 &  & +\left[(-iz)\{H_{\mathrm{int}},\; H_{\mathrm{aux}}\}_{b}+\frac{(-iz)^{2}}{2}\big\{\{H_{0},\; H_{\mathrm{aux}}\}_{b},\; H_{\mathrm{aux}}\big\}_{b}\right]\nonumber \\
 & & +\cdots \,,
\end{eqnarray}
where we have implied $H_{0}\equiv H_{0}(b,b^{*})$ and $H_{\mathrm{int}}\equiv H_{\mathrm{int}}(b,b^{*})$. 

By requiring that all nonresonant quartic terms of the new Hamiltonian $H'$ vanish, we can then determine the 
the coefficient functions $B_{1,2}$ of the auxiliary Hamiltonian $H_{\mathrm{aux}}$. This gives 
\begin{eqnarray}
\label{eq:solution-B}
&&B_{1}(\mathbf{k}_{1},\mathbf{k}_{2},\mathbf{k}_{3},\mathbf{k}_{4})  = \nonumber \\
&& \qquad i\frac{1}{z}\lambda\frac{(2\pi)^{3}}{\sqrt{16E_{k1}E_{k2}E_{k3}E_{k4}}(E_{k1}+E_{k2}+E_{k3}+E_{k4})}\,, \nonumber \\
&& B_{2}(\mathbf{k}_{1},\mathbf{k}_{2},\mathbf{k}_{3};\mathbf{k}_{4})  = \nonumber \\
&&  \qquad i\frac{1}{z}\lambda\frac{(2\pi)^{3}}{\sqrt{16E_{k1}E_{k2}E_{k3}E_{k4}}(E_{k1}+E_{k2}+E_{k3}-E_{k4})}\,. \nonumber \\
\end{eqnarray}
The derivation of the expression for the coefficients in Eq.~(\ref{eq:solution-B}) is given in Appendix \ref{sec:coef}.  With the coefficient functions of the auxiliary Hamiltonian determined according to Eq. (\ref{eq:solution-B}), the physical Hamiltonian in our new canonical variables $b_{\mathbf{p}}$ and $b_{\mathbf{p}}^{*}$ reads
\begin{eqnarray}
H^\prime & \approx & \int[d^{3}\mathbf{p}]E_{p}b_{\mathbf{p}}b_{\mathbf{p}}^{*}+\frac{\lambda}{16}\int\prod_{i=1}^{4}[d^{3}\mathbf{k}_{i}]\frac{1}{\sqrt{E_{k1}E_{k2}E_{k3}E_{k4}}}\nonumber \\
 &  & \times b_{\mathbf{k}1}b_{\mathbf{k}2}b_{\mathbf{k}3}^{*}b_{\mathbf{k}4}^{*}(2\pi)^{3}\delta(\mathbf{k}_{1}+\mathbf{k}_{2}-\mathbf{k}_{3}-\mathbf{k}_{4})\,,
\label{eq:ham-new}
\end{eqnarray}
where the first term has the form of a free Hamiltonian and the second term is an interaction term with only resonant contributions in these novel variables. We note that when transforming to these new variables,
according to Eq.~(\ref{eq:a-b-z}), terms of order $b_{\mathbf{p}}^{6}$, and of higher order, will emerge. In the next section, we will argue that all these terms are suppressed in the limit of large masses. 

With $H_{\mathrm{aux}}$ in Eq.~(\ref{eq:h-aux}), we then obtain from  Eq. (\ref{eq:a-b-z}) the 
explicit map from our original canonical variables $a_{\mathbf{p}}$ to our novel canonical variables  $b_{\mathbf{p}}$.
Each term on the r.h.s of Eq. (\ref{eq:a-b-z}) belongs to a specific order $\sim(\lambda\phi^{2}/m^{2})^{n}$ in coordinate
space. In order to obtain an inverse transformation of (\ref{eq:a-b-z}), or in other words to express $b_{\mathbf{p}}$ as a functional of $a_{\mathbf{p}}$,
we can equivalently rewrite Eq. (\ref{eq:a-b-z}) as 
\begin{eqnarray}
b_{\mathbf{p}} & = & a_{\mathbf{p}}+iz\{b_{\mathbf{p}},H_{\mathrm{aux}}(b,b^*)\}_{a} \nonumber \\
&& -\frac{(iz)^{2}}{2}\{\{b_{\mathbf{p}},H_{\mathrm{aux}}(b,b^*)\}_{a},H_{\mathrm{aux}}(b,b^*)\}_{a}+\cdots\,, \nonumber \\
\label{eq:a-b-z-1}
\end{eqnarray}
where we have used the fact that the Poisson brackets are invariant under a change of canonical basis. (See  Appendix \ref{sec:canonical} for a detailed discussion.) We note that $H_{\mathrm{aux}}$ in Eq.~(\ref{eq:h-aux}) is a functional of $b_{\mathbf{p}}$, while the
dependence of $b_{\mathbf{p}}$ on $a_{\mathbf{p}}$ (the creation-annihilation variables of the fundamental fields of the theory) is what we are searching 
for. However, Eq. (\ref{eq:a-b-z-1}) can always be solved perturbatively, where to leading order, $b_{\mathbf{p}}$
is just $a_{\mathbf{p}}$. At the next-to-leading order, we have to make the replacement $b_{\mathbf{p}}\rightarrow a_{\mathbf{p}}$
in the $iz\{b_{\mathbf{p}},H_{\mathrm{aux}}(b,b^*)\}_{a}$ term. 
Furthermore, at next-to-next-to-leading order, we have to include the $(1/2)(iz)^{2}\big\{\{b_{\mathbf{p}},H_{\mathrm{aux}}(b,b^*)\}_{a},H_{\mathrm{aux}}(b,b^*)\big\}_{a}$
term by replacing $b_{\mathbf{p}}\rightarrow a_{\mathbf{p}}$.  
Another contribution is the $z^2$ term from $iz\{b_{\mathbf{p}},H_{\mathrm{aux}}(b,b^*)\}_{a}$ 
where $b_{\mathbf{p}}$ is replaced by the next-to-leading order results, 
$b_{\mathbf{p}}\rightarrow a_{\mathbf{p}}+iz\{a_{\mathbf{p}},H_{\mathrm{aux}}(a,a^*)\}_{a}$. 
Thus in this fashion, one can obtain $b_{\mathbf{p}}$ and the Hamiltonian in Eq.~(\ref{eq:h-new}) as  functionals of $a_{\mathbf{p}}$ order by order in a systematic expansion. 

\section{Canonical mapping of $N=1$ scalars to Gross-Pitaevskii field theory}
\label{sec:mgp}

We will now employ the above formalism of canonical transformations to derive an effective nonrelativistic description of the infrared dynamics of massive scalar fields. We will derive explicitly that to leading order in the large mass limit $(p^2 \ll m^2$, $\lambda \phi^2 \ll  m^2)$, the infrared dynamics of the single component ($N=1$) $\lambda \phi^4$ theory can be described by the Gross-Pitaevskii (GP) equation. Before we begin, we briefly note that the GP equation
\begin{eqnarray}
\label{eq:GPEqnNR}
i \partial_{t} \psi(t,\mathbf{x}) = -\frac{\nabla^{2}}{2m} \psi(t,\mathbf{x})+g|\psi(t,\mathbf{x})|^{2}\psi(t,\mathbf{x})
\end{eqnarray}
was originally derived for the single particle wave-functions $\psi(t,\mathbf{x})$ of a many-body system of identical bosons with contact interactions with the scattering length $g m/(4\pi)$ employing a Hartree-Fock approximation~\cite{Gross:1961,Pitaevskii:1961}. Some of the recent interest in the GP equation derives from the fact that it provides a good model for the dynamics of cold atomic gases, including a variety of phenomena such as superfluidity and Bose-Einstein condensation \cite{Pethick:2002}. The derivation of the general properties of the GP equation has been widely discussed in the literature \cite{Floerchinger:1900mn,Schmitt:2014eka,Svistunov:2015}. 
For completeness, a brief discussion is provided in Appendix \ref{sec:dispersion}.  

We begin by defining the field $\psi(\mathbf{x})$ as the Fourier transform of the new canonical variables $b_{\mathbf{k}}$, 
\begin{equation}
\psi(\mathbf{x})=\int[d^{3}\mathbf{k}]\,b_{\mathbf{k}}\,e^{i\mathbf{k}\cdot\mathbf{x}}\,.
\label{eq:gp-field}
\end{equation}
We can express $\psi(\mathbf{x})$ as a functional of $\phi(\mathbf{x})$
and $\pi(\mathbf{x})$ by solving $b_{\mathbf{p}}$ as a functional of $a_{\mathbf{p}}$ by Eq. (\ref{eq:a-b-z-1}) 
and then expressing $a_{\mathbf{p}}$ in terms of $\phi _{\mathbf{p}}$ and $\pi _{\mathbf{p}}$ by using Eq. (\ref{eq:ak-ask}). 
In the large mass limit, the leading contribution in the expansion 
$\psi = \psi_{(0)}+\psi_{(1)}+\cdots$ coming from $a_{\mathbf{p}}$ in Eq. (\ref{eq:a-b-z-1}) is 
\begin{eqnarray}
\psi_{(0)}(\mathbf{x}) & = & \int [d^{3}\mathbf{k}] a_{\mathbf{k}}\,e^{i\mathbf{k}\cdot\mathbf{x}}\;,
\end{eqnarray}
which upon expanding to the order $O(\mathbf{k}^2/m^{2})$ can be evaluated as
\begin{eqnarray}
 \psi_{(0)}(\mathbf{x}) & = & \sqrt{\frac{m}{2}}\left(1-\frac{1}{4m^{2}}\nabla_{\mathbf{x}}^{2}\right)\phi(\mathbf{x}) \nonumber \\
 && +\frac{i}{\sqrt{2m}}\left(1+\frac{1}{4m^{2}}\nabla_{\mathbf{x}}^{2}\right)\pi(\mathbf{x})\;.
\end{eqnarray}
The above formula is consistent with Ref. \cite{Namjoo:2017nia} up to the time oscillation terms $e^{\pm imt}$. 
Introducing the short hand notation $\Phi\equiv\sqrt{m/2}\,\phi(\mathbf{x})$, $\Pi\equiv i\sqrt{1/(2m)}\,\pi(\mathbf{x})$ and henceforth denoting derivates as $O^{\prime}\equiv(\nabla_{\mathbf{x}}/m) O$ and similarly $O^{\prime\prime}\equiv(\nabla_{\mathbf{x}}^{2}/m^{2})O$, the leading contribution can be compactly expressed as 
\begin{eqnarray}
\label{eq:psi0}
\psi_{(0)}(\mathbf{x})& = & \Phi+\Pi-\frac{1}{4}\Phi^{\prime\prime}+\frac{1}{4}\Pi^{\prime\prime}\,.
\end{eqnarray}
The contribution from the second term in the expansion takes the form
\begin{eqnarray}
\psi_{(1)}(\mathbf{x}) & = & iz \int[d^{3}\mathbf{p}] e^{i\mathbf{p}\cdot\mathbf{x}}\{a_{\mathbf{p}},H_{\mathrm{aux}}(a,a^*)\}_{a}\,, \nonumber \\
\end{eqnarray}
and can be evaluated in the same way as
\begin{eqnarray}
\label{eq:psi1}
 \psi_{(1)}(\mathbf{x})  & =& -\frac{\lambda}{16m^{3}}\left[-\frac{5}{6}\Phi^{3}-\frac{7}{4}\Phi^{2}\Phi^{\prime\prime}-\frac{27}{8}\Phi\Phi^{\prime}\Phi^{\prime}+\frac{5}{2}\Phi^{2}\Pi\right.\nonumber \\
 &  & +\frac{11}{8}\Pi\Phi^{\prime}\Phi^{\prime}+\frac{3}{2}\Phi\Pi\Phi^{\prime\prime}+\frac{11}{4}\Phi\Phi^{\prime}\Pi^{\prime}+3\Phi^{2}\Pi^{\prime\prime}\nonumber \\
 &  & +\frac{3}{2}\Phi\Pi^{2}+\frac{5}{4}\Pi^{2}\Phi^{\prime\prime}+\frac{5}{4}\Pi\Phi^{\prime}\Pi^{\prime}+\frac{5}{8}\Phi\Pi^{\prime}\Pi^{\prime}\nonumber \\
 &  & +\left.2\Phi\Pi\Pi^{\prime\prime}-\frac{1}{2}\Pi^{3}-\frac{21}{8}\Pi\Pi^{\prime}\Pi^{\prime}-2\Pi^{2}\Pi^{\prime\prime}\right]\,.
\end{eqnarray}
A comparison of Eqs.~(\ref{eq:psi1}) and (\ref{eq:psi0}) shows that the ratio $\psi_{(1)}/\psi_{(0)}$ of the subleading to the leading term  is of order $\sim\lambda \phi^2/m^2$, corresponding to the ratio of the interaction energy $\lambda \phi^4$ to the mass density $m^2\phi^2$. Likewise, all higher order contributions $\psi_{(2)},\psi_{(3)},\cdots$, are suppressed by successive powers of $\lambda \phi^2/m^2$ and can therefore be neglected in the high mass limit where $\lambda \phi^2/m^2 \ll 1$ and $k^2\ll m^2$.

Similarly, one finds that in this nonrelativistic limit, quadratic and quartic terms in the Hamiltonian $H'(b,b^{*})$ dominate over higher order terms. To see
this, we express Eq. (\ref{H-prime}) in coordinate space and expand $E_{k}\approx m$ to leading order in $k^2\ll m^2$. One can then verify that the ratio $H^{\prime}(b_{\mathbf{p}}^{6})/H^{\prime}(b_{\mathbf{p}}^{4})$ is of order $O\left(\lambda\psi^{2}(\mathbf{x})/m^{3}\right)\sim O\left(\lambda\phi^{2}/m^{2}\right)$, consistent with the ratio of the quartic to quadratic term, $H(b_{\mathbf{p}}^{4})/H(b_{\mathbf{p}}^{2})\sim O\left(\lambda\phi^{2}/m^{2}\right)$. This power counting rule can be generalized to $H^{\prime}(b_{\mathbf{p}}^{2n+2})/H^{\prime}(b_{\mathbf{p}}^{2n})\sim O\left(\lambda\phi^{2}/m^{2}\right)$, suggesting that $\lambda\phi^{2}/m^{2}\ll1$ is a consistent expansion parameter in the large mass limit. 

We note that the limit $\lambda\phi^{2}/m^{2}\ll1$ does not necessarily mean nonrelativistic or small coupling constant separately; instead, this limit implies that the interaction energy is much smaller than the kinetic energy in the $\lambda\phi^{4}$ theory. Another important point is that the dynamics of the system we are interested in is classical which suggests that the occupation number for a particular
momentum mode must be much larger than the quantum one-half occupancy per mode. This is equivalent to the condition $|b_{\mathbf{k}}|^{2}/V\sim V|\psi|^{2}\sim mV\phi^{2}\gg1$. 
In other words, application of the classical theory is justified by the large particle number in the superfluid and it is independent of the limit $\lambda\phi^{2}/m^{2}\ll1$. 


Following our discussion of the power counting, the Hamiltonian in our new canonical variables in the limit $\lambda\phi^{2}/m^{2}\ll1$  and $\mathbf{k}^2\ll m^2$ reads 
\begin{eqnarray}
& &H'  \approx  m\int[d^{3}\mathbf{p}]b_{\mathbf{p}}b_{\mathbf{p}}^{*}+\int[d^{3}\mathbf{p}]\frac{\mathbf{p}^{2}}{2m}b_{\mathbf{p}}b_{\mathbf{p}}^{*}\nonumber \\
 &  & +\frac{\lambda}{16m^{2}}\int\prod_{i=1}^{4}[d^{3}\mathbf{k}_{i}]b_{\mathbf{k}1}b_{\mathbf{k}2}b_{\mathbf{k}3}^{*}b_{\mathbf{k}4}^{*} \nonumber \\
& & \qquad \qquad (2\pi)^{3}\delta(\mathbf{k}_{1}+\mathbf{k}_{2}-\mathbf{k}_{3}-\mathbf{k}_{4})\;.\nonumber \\
 \end{eqnarray}
Expressed in terms of the coordinate space fields, this effective infrared Hamiltonian takes then takes the form
 \begin{eqnarray}
 H' & = & m\int d^{3}\mathbf{x}\,|\psi(t,\mathbf{x})|^{2}+\int d^{3}\mathbf{x}\,\psi^{*}(t,\mathbf{x})\left(-\frac{\nabla^{2}}{2m}\right)\psi(t,\mathbf{x})\nonumber \\
 &  & +\frac{\lambda}{16m^{2}}\int d^{3}\mathbf{x}\,|\psi(t,\mathbf{x})|^{4}\,,
\label{eq:ham-gp}
\end{eqnarray}
which following a straightforward set of manipulations gives rise to the following equation of motion
\begin{eqnarray}
i\frac{\partial}{\partial t}\psi(t,\mathbf{x}) & = & \left(m-\frac{\nabla^{2}}{2m} \right)\psi(t,\mathbf{x})+\frac{\lambda}{8m^{2}}|\psi(t,\mathbf{x})|^{2}\psi(t,\mathbf{x}). \nonumber \\
\label{eq:eom}
\end{eqnarray}
where we have now explicitly distinguished between the spatial and time dependence of $\psi(\mathbf{x})$. Except for the $m \psi(t,\mathbf{x})$ term on the RHS, which can be absorbed into a redefinition of the complex scalar field $\psi(t,\mathbf{x}) \to e^{imt}\psi(t,\mathbf{x})$, this is identical with the GP equation in (\ref{eq:GPEqnNR}) with the coupling constant $g$ identified as $g=\lambda/(8m^{2})$. While there has been a prior qualitative discussion of the Gross-Pitaevski equation as an effective description of the infrared dynamics of massive relativistic scalar fields~\cite{Berges:2014xea}, our derivation provides an explicit map between the relativistic and nonrelativistic descriptions (given in Eqs.~(\ref{eq:psi1}) and (\ref{eq:psi0})) and enables systematic extensions of the correspondence to higher orders.


\begin{figure*}[t!]
\centering
\includegraphics[width=0.45\textwidth]{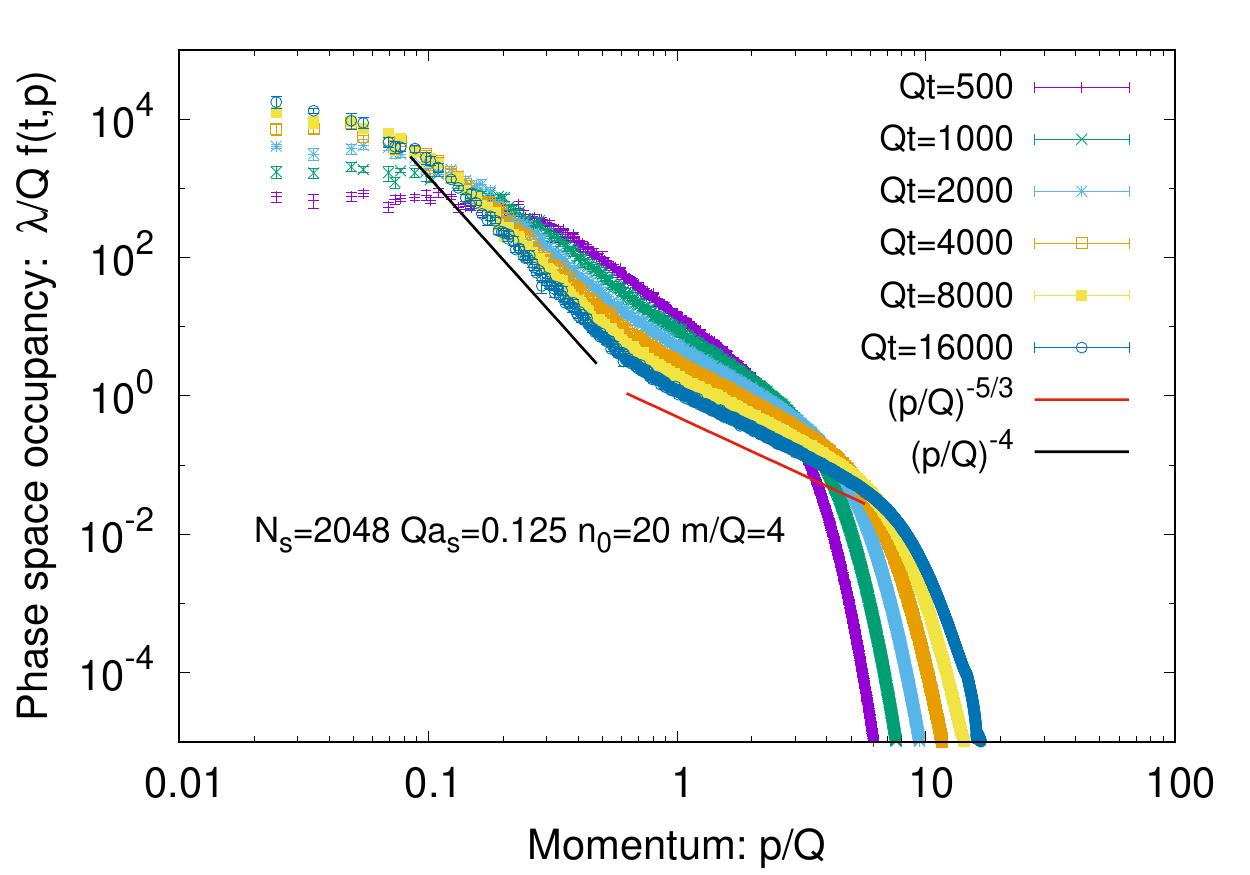}
\includegraphics[width=0.45\textwidth]{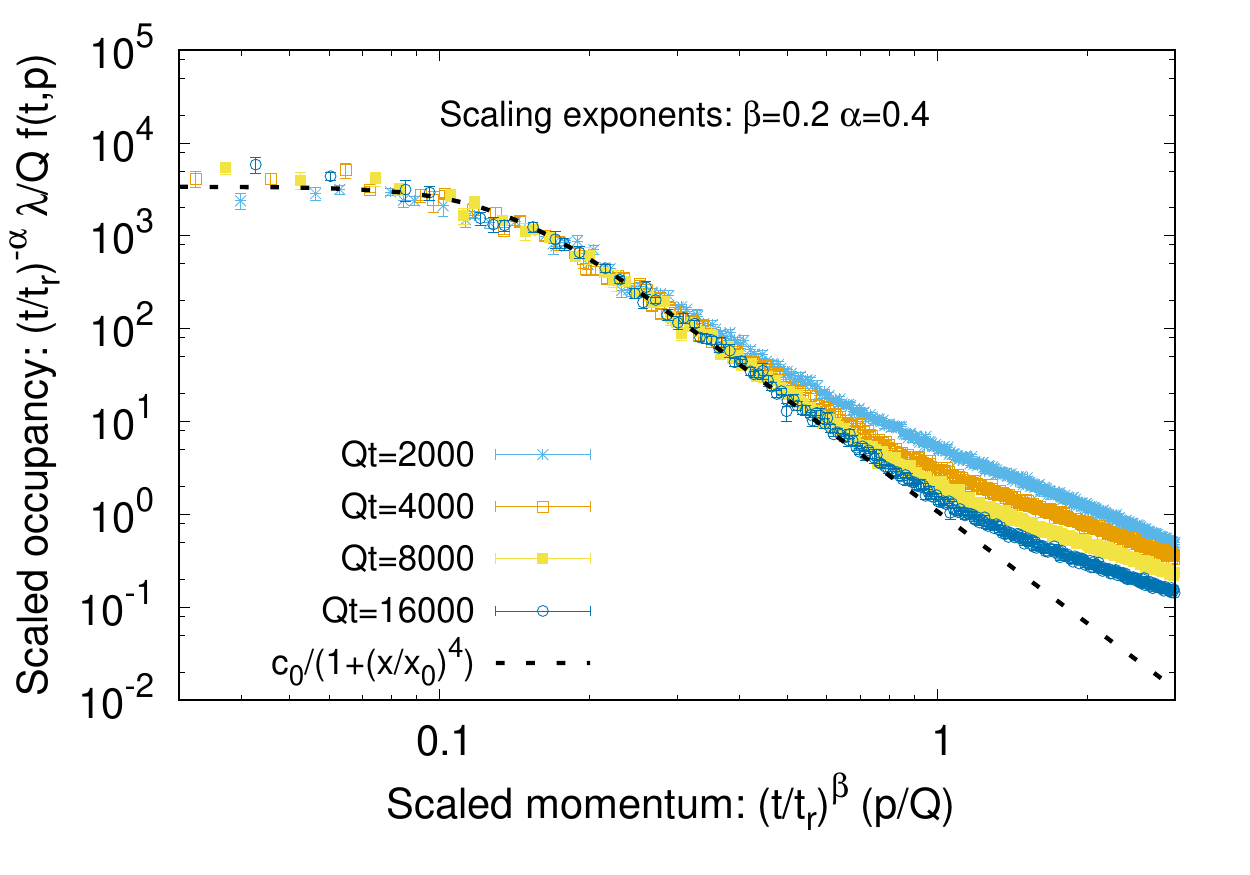}
\includegraphics[width=0.45\textwidth]{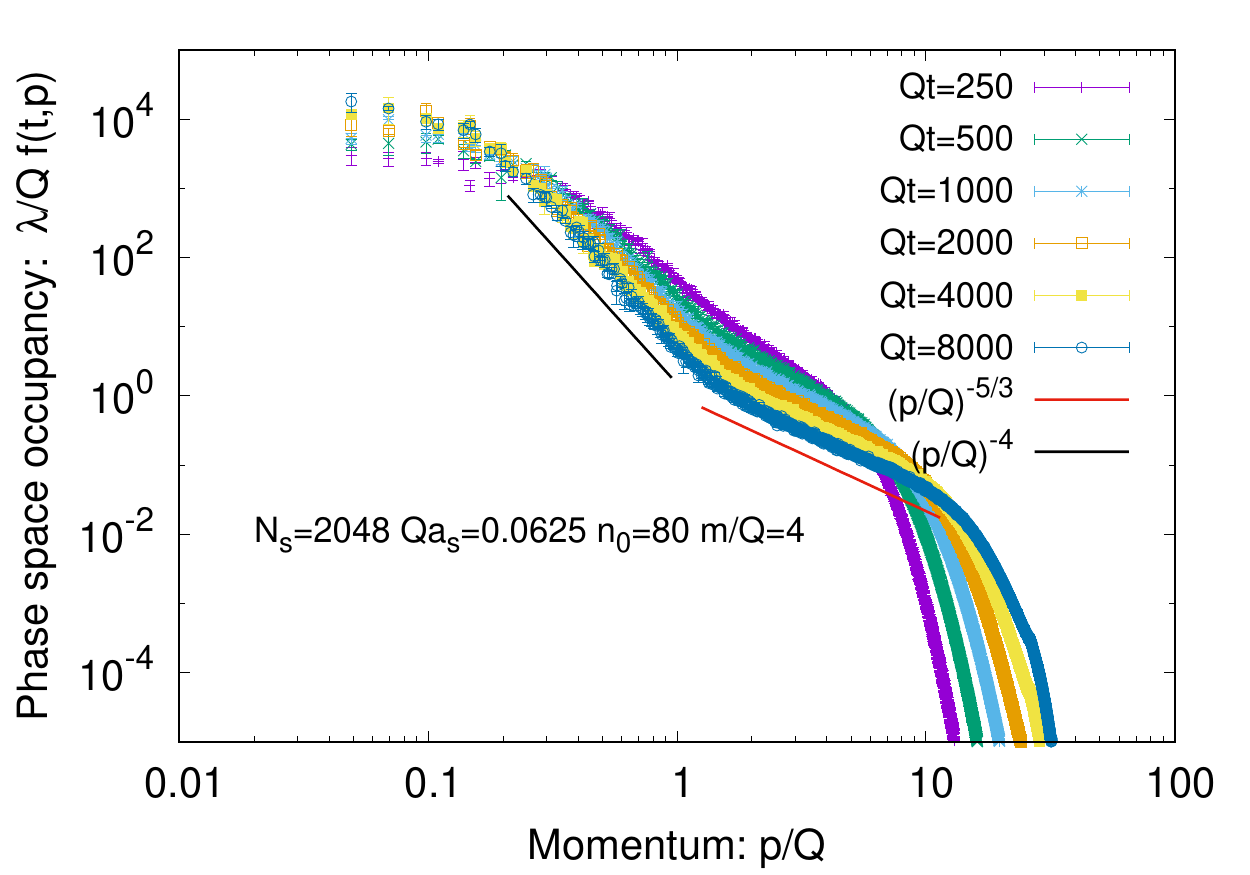}
\includegraphics[width=0.45\textwidth]{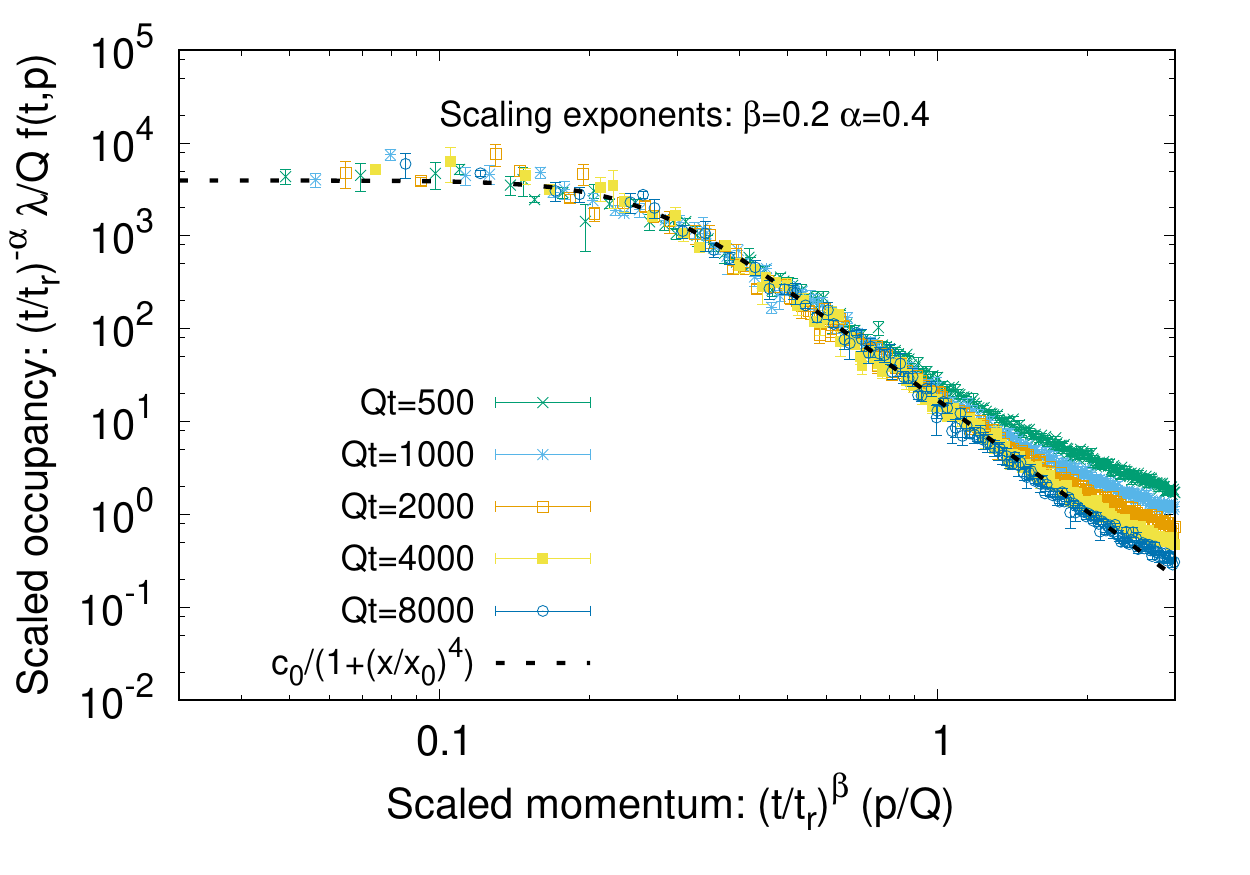}
\caption{(left) Evolution of the single particle spectrum of the relativistic scalar theory for characteristic values of the parameters. (right) Self-similar scaling of the (inverse) infrared cascade leads to a collapse of the data onto a universal scaling curve. Data shown in the top panel is for $m/Q=4$ and $n_0=20$ while data in the bottom panels is for $m/Q=4$ and $n_0=80$ to maximize the scaling window. }
\label{fig-spectra}       
\end{figure*}

\section{Numerical simulations of the dynamics of overoccupied massive relativistic scalar fields}
\label{numerical}
We will now illustrate the power of this approach to understand nonequilibrium phenomena of relativistic scalar fields in terms of the effective nonrelativistic infrared degrees of freedom. Our simulations are performed for massive relativistic scalar fields in $D=2$ spatial dimensions by solving the classical field equations of motion
\begin{eqnarray}
\partial_{t} \phi(t,\xt) &=& \pi(t,\xt)\;, \nonumber \\
\partial_{t} \pi(t,\xt) &=& \partial_{i}\partial^{i} \phi(t,\xt) -m^2 \phi(t,\xt) -\frac{\lambda}{6}  \phi(t,\xt)^3\;,
\nonumber\\
\end{eqnarray}
where $\partial_{i}\partial^{i}$ is the usual Laplacian. We discretize the theory on a $N_s \times N_s$ spatial lattice with spacing $a_s$ and using a leap-frog scheme with time step $a_t/a_s=0.05$. We choose in the simulations lattice sizes from $512^2$ up to $2048^2$ with lattice spacings $a_s=(0.0625-0.125)~Q^{-1}$, where $Q$ is a characteristic momentum scale of the initial condition. Since the classical equations of motion admit a rescaling of the form $\phi \to \lambda^{1/2} \phi\;,~\pi \to \lambda^{1/2} \pi$, the coupling constant $\lambda$ effectively drops out of the classical equations of motion and only enters via the initial condition.

We can simulate the far from equilibrium dynamics starting from the initial conditions for the fields given by a random superposition of free field modes,
\begin{eqnarray}
\phi_{0}(\xt)=\frac{1}{(N_s a_s)^2}\sum_{\mathbf{p}}\frac{1}{\sqrt{2E_{p}}} \Big[\alpha_{\mathbf{p}} e^{+i\mathbf{p}\cdot\xt} + \alpha^{*}_{\mathbf{p}} e^{-i\mathbf{p}\cdot\xt}  \Big]\;, \nonumber \\
\pi_{0}(\xt)=\frac{(-i)}{(N_s a_s)^2} \sum_{\mathbf{p}}\sqrt{\frac{E_{p}}{2}}  \Big[\alpha_{\mathbf{p}} e^{+i\mathbf{p}\cdot\xt} - \alpha^{*}_{\mathbf{p}} e^{-i\mathbf{p}\cdot\xt}  \Big]\;, \nonumber \\
\end{eqnarray}
where $\alpha_{\mathbf{p}}$ and $\alpha^{*}_{\mathbf{p}}$ are sampled with Gaussian magnitude 
and uniform random phase distribution, such that
\begin{eqnarray}
\langle \alpha_{\mathbf{p}} \alpha^{*}_{\mathbf{q}} \rangle = (N_s a_s)^2 \delta_{\mathbf{p},\mathbf{q}} 
f(t=0,p)
\;,
\end{eqnarray}
with the initial phase space distribution or the occupancy per unit of phase space volume
\begin{eqnarray}
f(t=0,p)=\frac{6 n_0 Q}{\lambda} \theta\Big(Q-p\Big)\;.
\end{eqnarray}
Here $\lambda/Q$ is the dimensionless coupling, $Q$ is the dimensionful scale characterizing the initial momentum and the dimensionless parameter $n_0$ characterizes the degree of initial overoccupancy.

\subsection{Nonthermal fixed point in massive relativistic scalar theory}
\label{sec:spectra-ntfp}
We first analyze the dynamics of the relativistic fields by calculating the single-particle spectra of the relativistic field  according to the standard definition~\cite{Berges:2015kfa}
\begin{equation}
f(t,p)
= \frac{1}{(N_s a_s)^2} \sqrt{ \Big\langle |\tilde{\phi}(t,\mathbf{p})|^2 \Big\rangle \Big\langle |\tilde{\pi}(t,\mathbf{p})|^2 \Big\rangle}\,,
\end{equation}
where $\tilde{\phi}(t,\mathbf{p})=a_s^2\sum_{\xt}~\phi(t,\xt)~e^{-i\mathbf{p}\cdot\xt}$ and $\tilde{\pi}(t,\mathbf{p})=a_s^2\sum_{\xt}~\pi(t,\xt)~e^{-i\mathbf{p}\cdot\xt}$ denote the Fourier transform of the relativistic field and that of its conjugate momentum field respectively. 

Our results are depicted in Fig.~\ref{fig-spectra}, where in the left panel we show the time evolution of the single particle spectrum as a function of momentum for $m/Q=4$ and $n_0=20$ in the top panel and $m/Q=4$ and $n_0=80$ in the bottom panel. One observes that after a short transient dynamics, whose duration depends sensitively on the initial overoccupancy $n_0$, the spectrum approaches a nonthermal fixed point with a quasi-stationary spectrum characterized by a bimodal power law. These power law dependencies are indicated by the black/red lines in Fig.~\ref{fig-spectra}. As noted previously~\cite{Berges:2012us,Berges:2013lsa,Orioli:2015dxa}, this behavior can be interpreted in terms of a dual cascade, corresponding to a simultaneous flux of energy towards the ultraviolet and a flux of particle number towards the infrared.

While the dynamics of the ultraviolet (energy) cascade can be understood in the framework of (perturbative) kinetic theory~\cite{Micha:2004bv,Berges:2013fga,Berges:2013eia}, the infrared cascade features large occupation numbers $(\lambda/Q) f \gg 1$ and is therefore intrinsically nonperturbative in nature. However, recent analyses \cite{Orioli:2015dxa,Walz:2017ffj} suggest that important features of such nonperturbative infrared behavior can be understood in terms of a vertex resummed kinetic description based on a 2PI 1/$N$  expansion to next-to-leading order for $N$ component scalar fields. These analyses suggest that one should expect a self-similar evolution of the infrared occupancy described by a statistical single particle distribution satisfying the relation, 
\begin{eqnarray}
\label{eq:ss_scaling}
f(t,p)=(t/t_r)^{\alpha} f_{S}((t/t_r)^{\beta} p)\,,
\end{eqnarray}
with scaling exponents $\alpha=1$ and $\beta=1/2$ in $D=2$ dimensions, and the reference time scale $t_r$. We note that the scaling relation $\alpha= D \beta$ follows directly from the constraint of particle number conservation, $\int d^Dp~f(t,p)={\rm const.}$, in the infrared. Conversely, the scaling exponent $\beta$ reflects the underlying microscopic dynamics of the system and needs to be determined from a scaling analysis of the relevant dynamical processes~\cite{Orioli:2015dxa}.

\begin{figure*}[t!]
\centering
\includegraphics[width=0.45\textwidth]{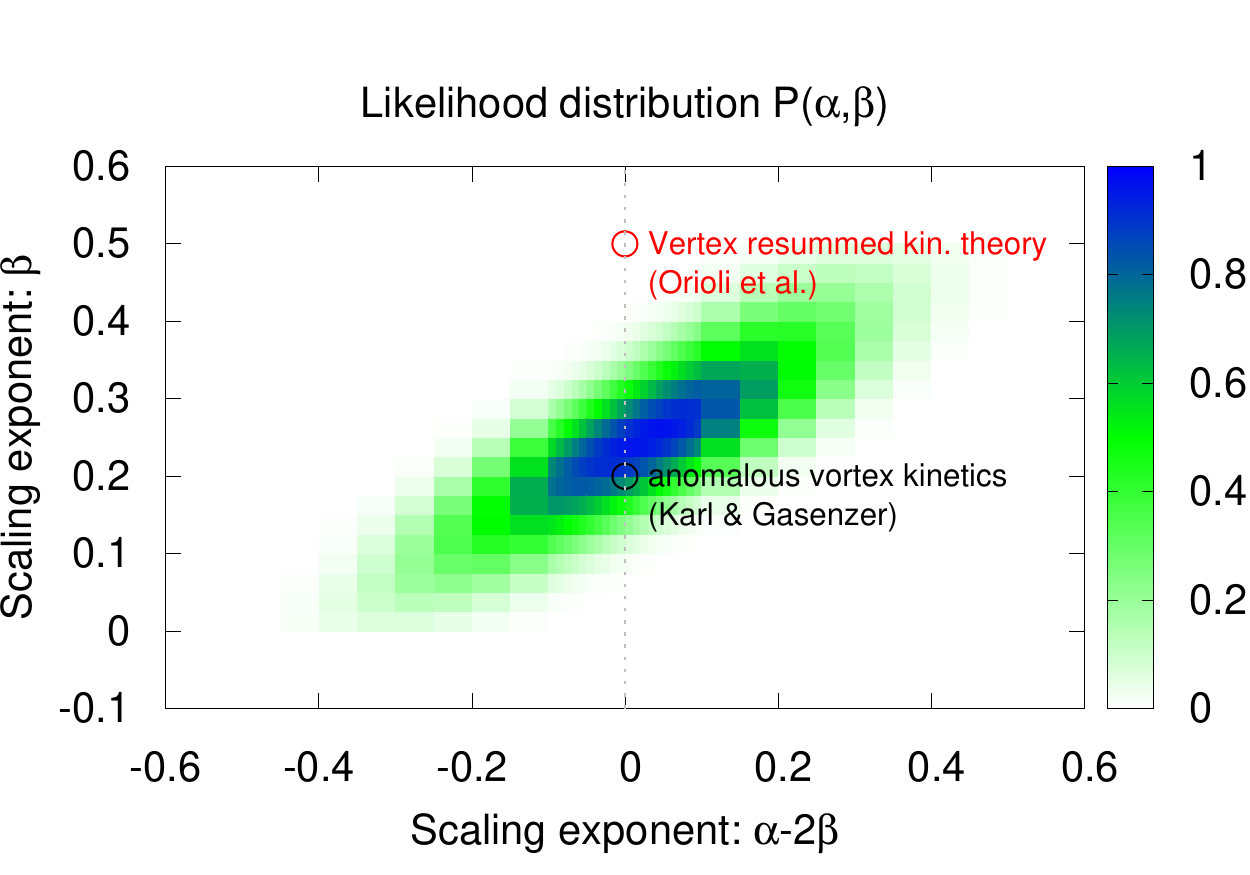}
\includegraphics[width=0.45\textwidth]{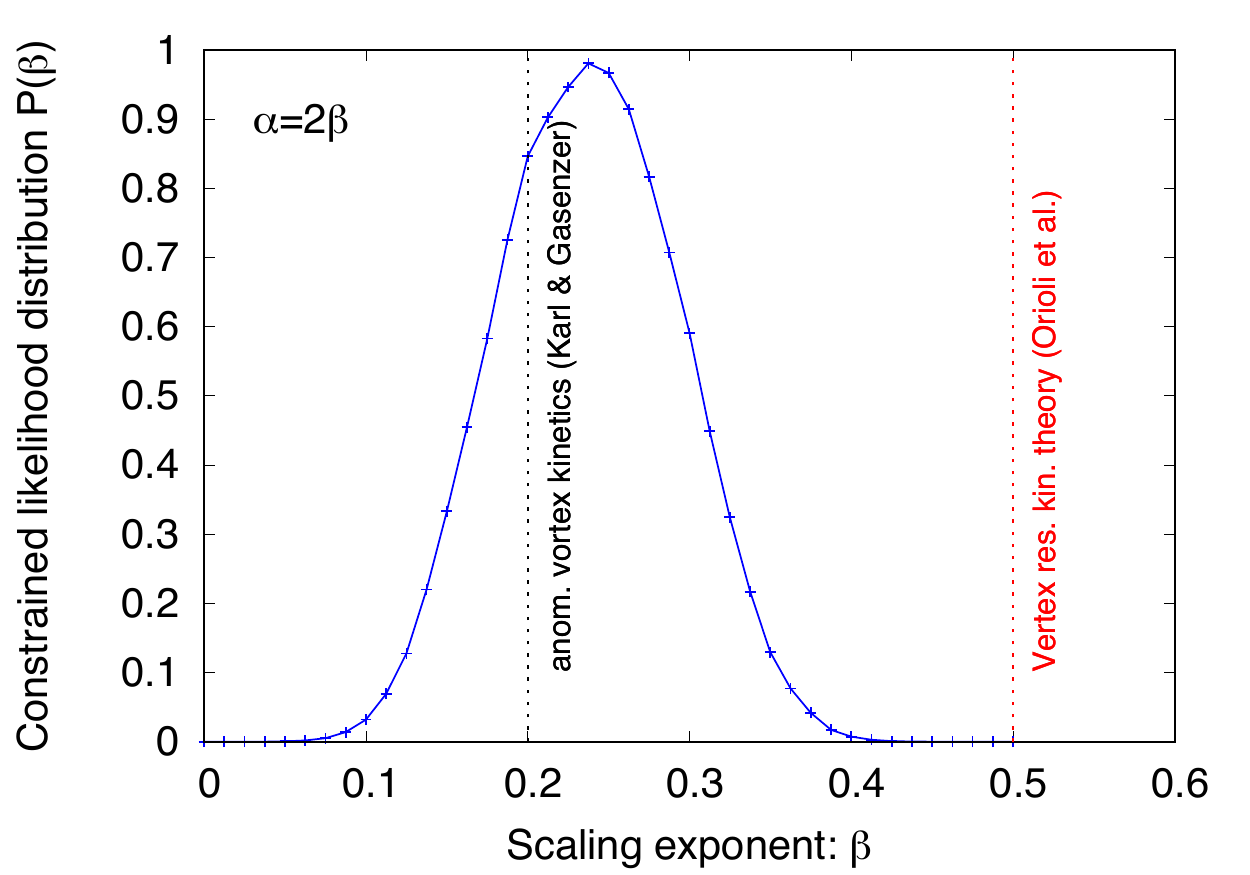}
\caption{(left) Likelihood distribution of scaling exponents. (right) Constraint likelihood distribution under the particle number conservation constraint  ($\alpha=2\beta$). Comparisons to predicted values of the exponents $\alpha,\beta$ from vertex resummed kinetic theory \cite{Orioli:2015dxa} (red) and the strongly anomalous fixed point of a nonrelativistic Bose Gas \cite{Karl:2016wko} (black) are also shown.}
\label{fig-chisqr}       
\end{figure*}

Since Eq.~(\ref{eq:ss_scaling}) predicts that all spectra at different times (under appropriate rescaling of momenta and phase space occupancy) collapse onto a universal scaling curve $f_{S}(x)$~\cite{Berges:2013fga,Berges:2013eia,Orioli:2015dxa,Schlichting:2012es}, it is straightforward to check whether such behavior is observed in our simulations. Our results for the rescaled spectra are depicted in the right panel of Fig.~\ref{fig-spectra}, where we show the rescaled phase space occupancy $(t/t_r)^{-\alpha} f$ as a function of the rescaled momentum variable $(t/t_r)^{\beta} p$. We see strikingly that the data lie on a universal curve which is well approximated by the functional form
\begin{eqnarray}
f_{S}(x)\simeq \frac{c_0}{1+(x/x_0)^{4}}\;,
\end{eqnarray}
thereby providing strong evidence of self-similarity. We further observe that the associated scaling exponents $\alpha$ and $\beta$ are significantly different from the values expected from the aforementioned 2PI 1/$N$ analysis. In order to quantify this discrepancy, we performed a statistical analysis of the self-similarity of moments $f^{(m)}(t,p)=p^{m} f(t,p)$ of the distribution. From the scaling hypothesis in Eq.~(\ref{eq:ss_scaling}), one can derive the following scaling relation of the moments
\begin{eqnarray}
\label{eq:ScalingM2}
\left(\frac{t}{t_r}\right)^{-\alpha+m\beta} f^{(m)}\left(t,\left(\frac{t}{t_r}\right)^{-\beta} p\right)=p^{m} f(t,p)\;.  \nonumber \\
\end{eqnarray}
Defining the LHS of Eq.~(\ref{eq:ScalingM2}) as $M^{(m)}_{\alpha,\beta,t_r}(t, p)$ and following Ref. \cite{Berges:2013fga}, this scaling relation can be used to quantify the deviation from perfect scaling behavior by computing the chi-squared function:
\begin{eqnarray}
&&\chi^{2}(\alpha,\beta)= \nonumber\\
&&\int_{p_{\mathrm{min}}}^{p_{\mathrm{max}}} d p~\frac{ \sum_{i,j} \left[ M^{(2)}_{\alpha,\beta,t_r}(t_i, p) - M^{(2)}_{\alpha,\beta,t_r}(t_j, p)\right]  ^2 } {\sum_{i}  \left[ M^{(2)}_{\alpha,\beta,t_r}(t_i, p) \right]^2}\;, \nonumber\\
\end{eqnarray}
for momenta $p_{\mathrm{min}} < p < p_{\mathrm{max}}$ in the scaling regime. Our results of the analysis for $p_{\mathrm{min}}/Q=0.08$ and $p_{\mathrm{max}}=0.8$ between times $Q t_{i} \in \{1000,2000,3000,4000,8000 \}$ (with $Q t_r=3000$) from the $m/Q=4$, $n_0=80$ data set, are presented in Fig.~\ref{fig-chisqr}, where we show the likelihood distribution
\begin{eqnarray}
P(\alpha,\beta) \sim \exp\left[ -\frac{\chi^{2}(\alpha,\beta)}{ 2\chi_{\mathrm{min}}^{2}}\right]\;.
\end{eqnarray}
Here $\chi_{\mathrm{min}}^{2}=\mathrm{min}_{\alpha,\beta} \chi^{2}(\alpha,\beta)$ corresponds to the best convergence of the rescaled data to a universal curve. Despite the fact that the extraction of the exponents $\beta$ and $\alpha-2\beta$ is strongly correlated\footnote{Note that this correlation can be attributed in part to the fact that the self-similar scaling of the $\sim(p/Q)^{-4}$ power law tail is only sensitive to the linear combination of exponents $\alpha-4\beta$.}, one observes that the analysis strongly favors values of $\beta$ between $0.1\lesssim \beta \lesssim 0.35$. By implementing the particle number conservation constraint explicitly, we obtain the constrained likelihood distribution shown in the right panel of Fig.~\ref{fig-chisqr}, from which we infer the following estimate of the scaling exponent
\begin{eqnarray}
\beta = 0.24 \pm 0.08\;.
\label{eq:betaval}
\end{eqnarray}
While the observed values of $\beta$ are significantly different from the 2PI $1/N$ prediction ($\beta=1/2$ shown in red) \cite{Orioli:2015dxa}, we find that our results are in good agreement with the values reported in \cite{Karl:2016wko} for a ``strongly anomalous'' nonthermal fixed point in a nonrelativistic single component Bose gas ($\beta=1/5$, shown in black) described by the GP equation. The arguments of  \cite{Karl:2016wko} attribute this value of $\beta$ to the underlying dynamics of vortex defects. Specifically, this dynamics corresponds to an increase of the mean distance between defects arising from the dilution in the number of vortices as a consequence of three-body vortex-vortex-antivortex collisions. Even though it is by no means obvious that similar processes could emerge in the deep infrared sector of the massive $\lambda \phi^4$ theory, this analysis clearly suggests that a more careful analysis of the infrared dynamics in terms of the effectively nonrelativistic infrared degrees of freedom is required to understand the observed scaling behavior.

\subsection{Dynamics of nonrelativistic infrared modes}
\begin{figure}[t!]
\includegraphics[width=\textwidth]{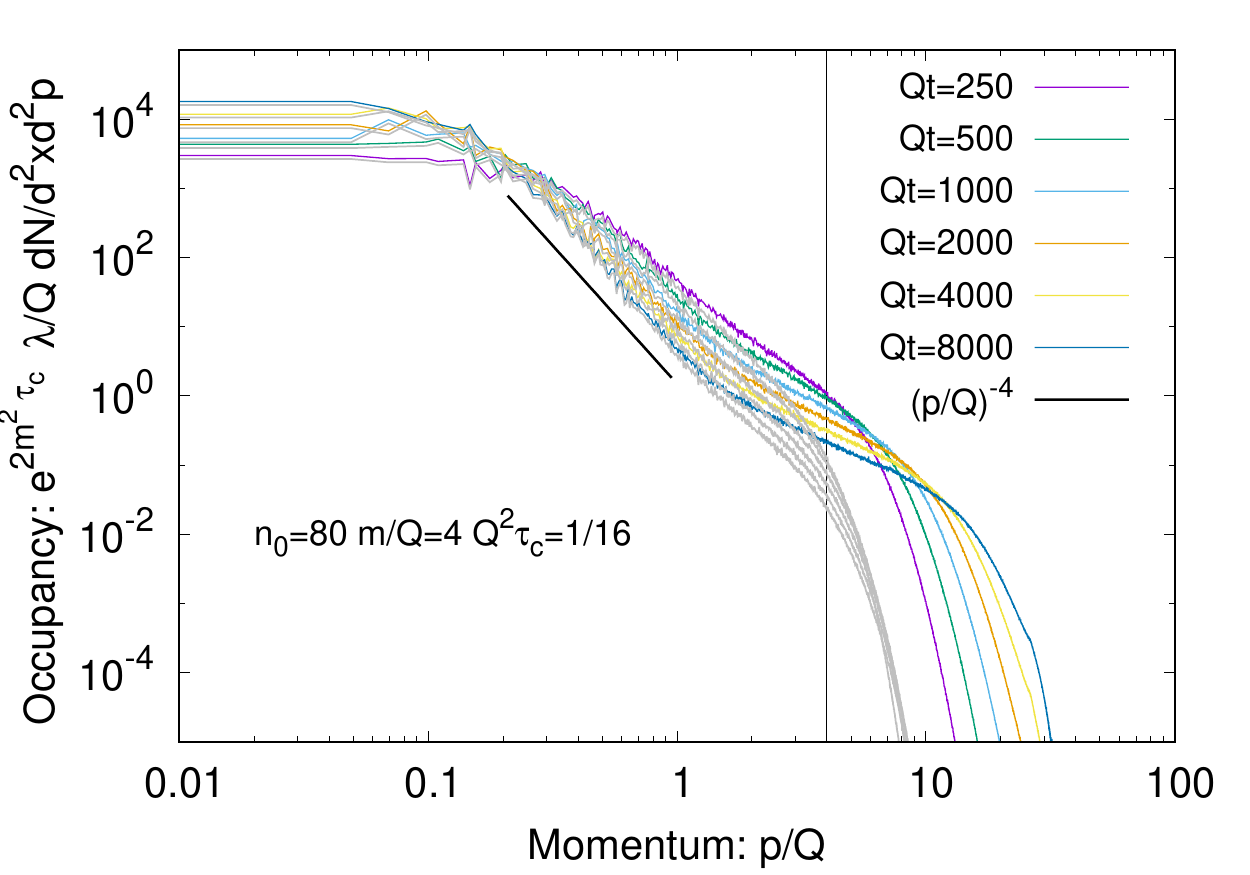}
\caption{Spectra of the phase-space occupancy before (colored lines) and after (gray lines) gradient flow cooling up to $Q^2 \tau_{c}=1/16$. Cooling effectively removes excitations beyond the mass scale, indicated by the vertical black line.}
\label{fig-Cooling}    
\end{figure}

We will now investigate to what extent the phenomena described above can be understood in terms of nonrelativistic degrees of freedom that emerge in the infrared. We expect these degrees of freedom to be distinct for long wavelength modes with $p\ll m$. However since at late times the relativistic scalar field theory contains excitations with momenta $p \gg m$, we have to first isolate the long wavelength excitations of interest. In practice, this is achieved by employing a gradient flow cooling technique, where the scalar field configurations $\phi(t,\tau_c=0,\xt)=\phi(t,\xt)$ are evolved according to
\begin{eqnarray}
&& \partial_{\tau_c} \phi(t,\tau_c,\xt)=   \nonumber\\
 && \qquad \qquad \partial_{i}\partial^{i} \phi(t,\tau_c,\xt) -m^2 \phi(t,\tau_c,\xt) -\frac{\lambda}{6}  \phi(t,\tau_c,\xt)^3\;. \nonumber\\
\end{eqnarray}
and cooled momentum fields $\pi(t,\tau_c,\xt)$ are constructed from scalar fields at adjacent time steps as 
$\pi(t,\tau_c,\xt)=[\phi(t+\Delta t/2,\tau_c,\xt)-\phi(t-\Delta t/2,\tau_c,\xt)]/\Delta t$. 
Since the cooling time $\tau_c$ is a measure of the extent to which ultraviolet modes have been removed from the simulation results, a choice of $m^2 \tau_c \sim 1$ ensures that excitations with momenta larger than the mass have been efficiently removed as shown explicitly in Fig.~\ref{fig-Cooling}. We note that for sufficiently large mass, the cooling also suppresses low energy excitations by a factor of $\sim \exp(-m^2 \tau_c)$. We will account for this suppression, by rescaling the cooled fields by the inverse of this factor when comparing results obtained from cooled configurations with ones obtained directly from the relativistic scalar fields.

%

\begin{figure}[t!]
\includegraphics[width=\textwidth]{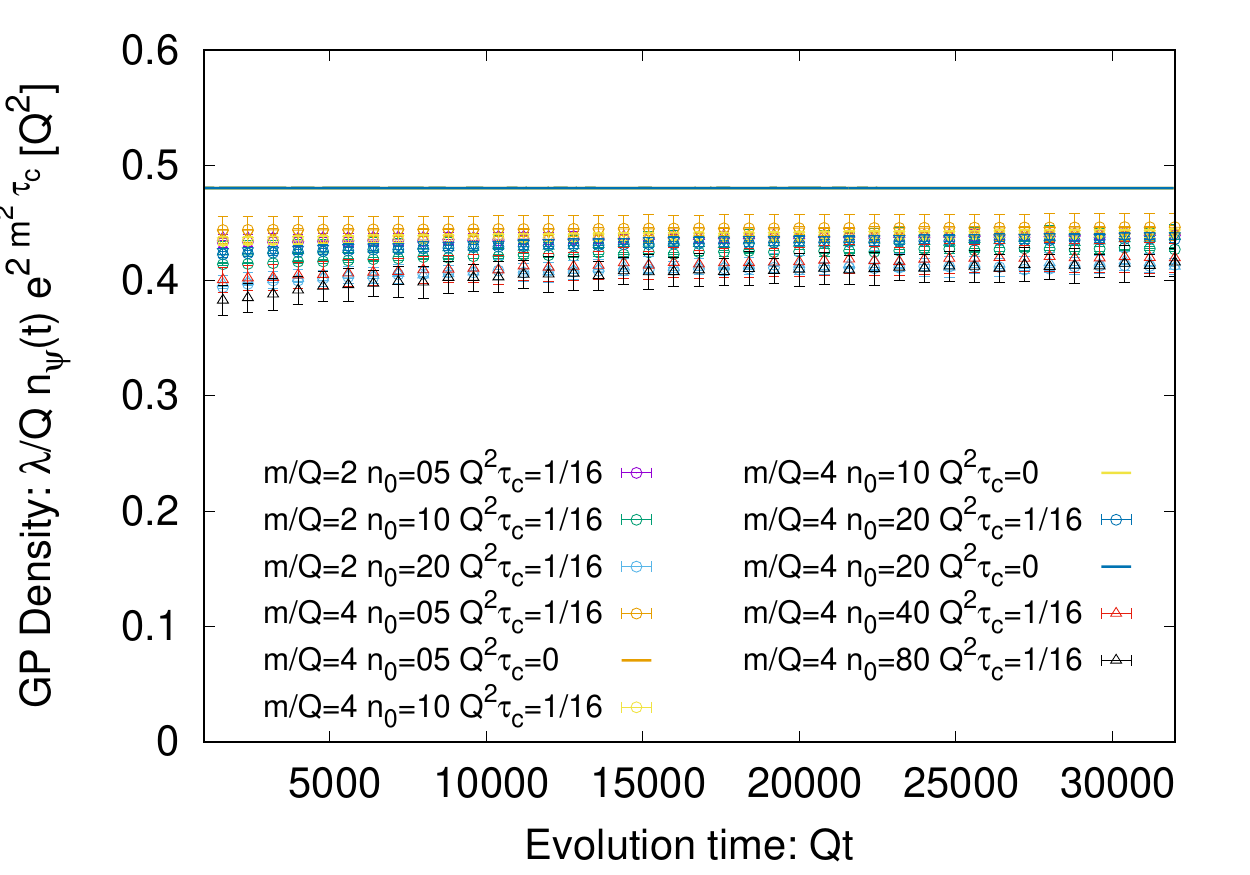}
\caption{Evolution of the particle number density $n_{\psi}(t)$ of the Gross-Pitaevskii field. Symbols correspond to a reconstruction of the GP field from cooled field configurations $(Q^2\tau_{c}=1/16$). Solid lines (that fully overlap) depict the GP density obtained directly from the relativistic scalar field without cooling.}
\label{fig-GrossPitaevskiDensity}    
\end{figure}

With the cooled configurations in hand, we can utilize the mapping spelled out in Eqs.~(\ref{eq:psi0}) and (\ref{eq:psi1}) to construct the effective nonrelativistic field $\psi(t,\mathbf{x})$ at each point of the space-time evolution. As we discussed previously, we anticipate the dynamical evolution of this field to be described approximately by the GP equation albeit one should also anticipate a coupling of this field to the hard modes in the system. We find that on average the number density of nonrelativistic modes
\begin{eqnarray}
n_{\psi}(t)=\frac{1}{N_s^2} \sum_{\mathbf{x}} |\psi(t,\mathbf{x})|^2\,,
\end{eqnarray}
is approximately conserved. This is seen in Fig.~\ref{fig-GrossPitaevskiDensity}, where we present results for different values of the mass ratio $m/Q$ and different degrees of initial occupancy $n_0$. Scaling of the curves by a factor of $n_{0}^{-1}$, leads to similar results for different values of $n_0$, where in all cases, the number density is approximately conserved up to the few percent level. Even though the nonrelativistic infrared excitations are in principle coupled to the hard  ultraviolet modes, our result shows that after a short transient period the coupling between soft and hard modes appears to be weak. Therefore the dynamics of the infrared sector can be studied separately from the ultraviolet cascade.

\begin{table}[b!]
\begin{tabular}{|c|c|c|c||c|c|c|c|}
\hline
$m/Q$ & $n_0$ & $(\lambda /Q) n_{\psi} [Q^2]$ & $Q\xi _h$ & $m/Q$ & $n_0$ & $(\lambda /Q)n_{\psi} [Q^2]$ & $Q\xi _h$\\ \hline
2 & 5 & 2.2 & 1.91 & 4 & 10 & 4.5  & 1.88 \\ \hline
2 & 10 & 4.4 & 1.34 & 4 & 20 & 9.0 & 1.33 \\ \hline
2 & 20 & 8.4 & 0.97  & 4 & 40 & 17.2 & 0.96 \\ \hline
4 & 5 & 2.3 & 2.63 & 4 & 80 & 33.6 & 0.69 \\ \hline
\end{tabular}
\caption{Values of the number density $n_{\psi}$ and the effective vortex healing length $\xi_{h}$ extracted for different simulation parameters. Note that we have included the cooling factor $e^{2m^2\tau_c}$ into the listed values of $(\lambda /Q) n_{\psi} [Q^2]$.}
\label{tab-hl}
\end{table}

\begin{figure*}[tp!]
\centering
\includegraphics[width=0.45\textwidth]{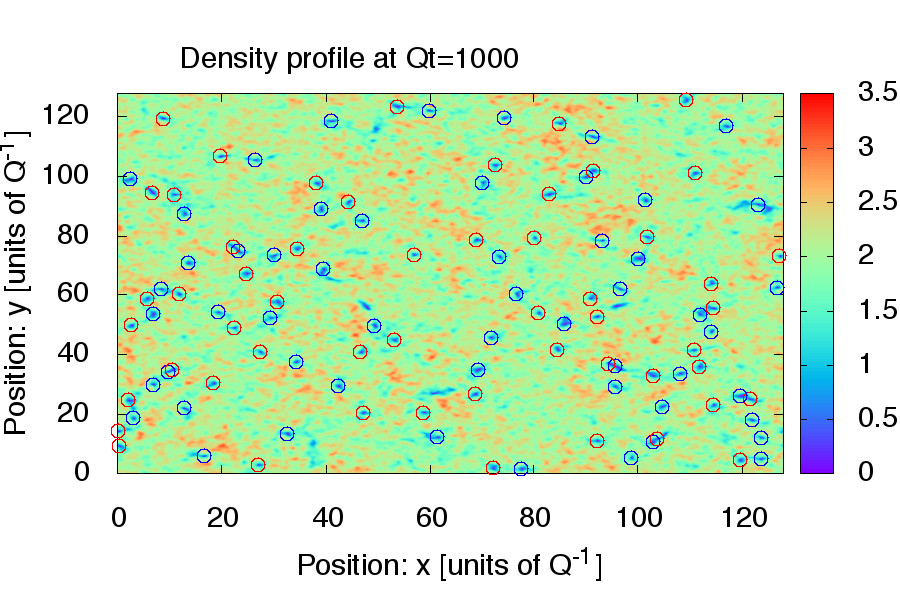}
\includegraphics[width=0.45\textwidth]{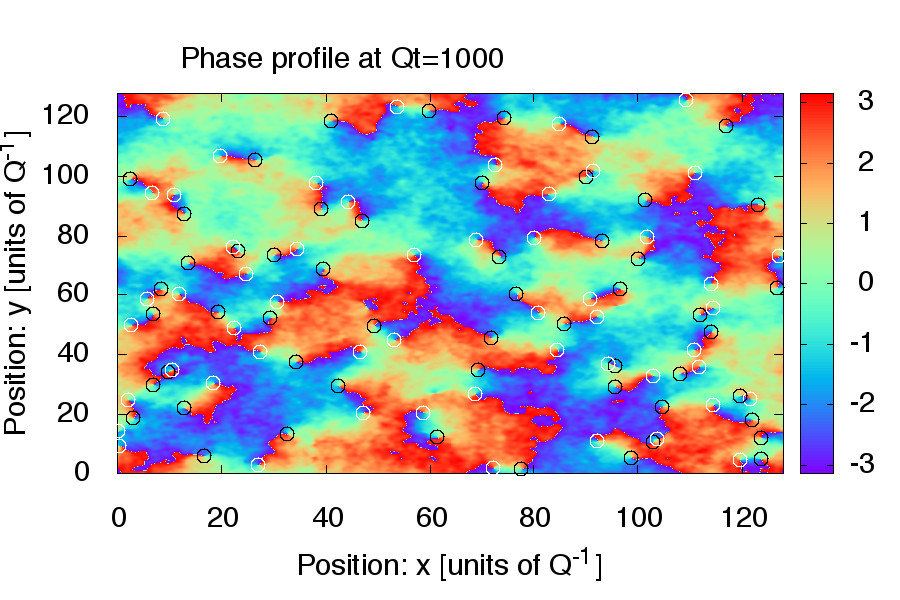}
\includegraphics[width=0.45\textwidth]{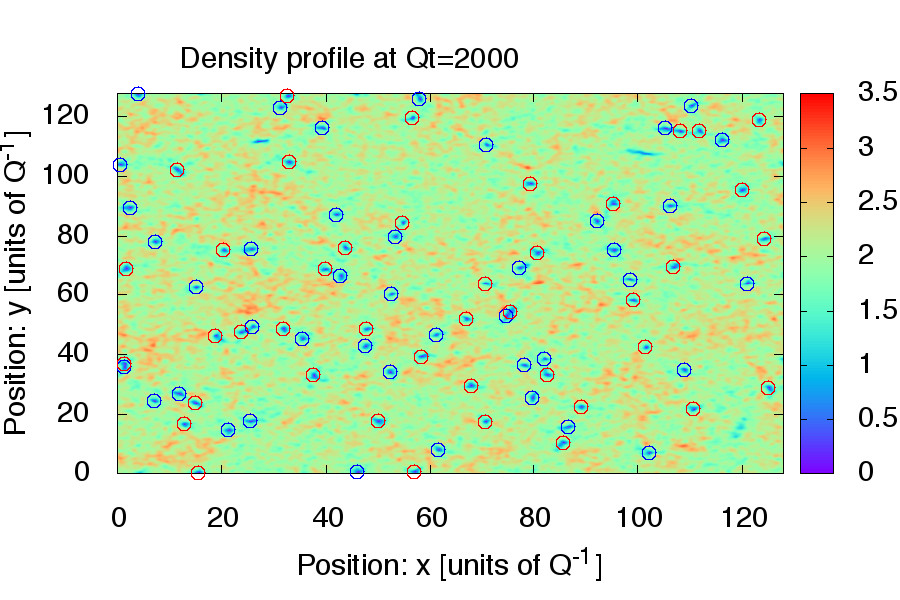}
\includegraphics[width=0.45\textwidth]{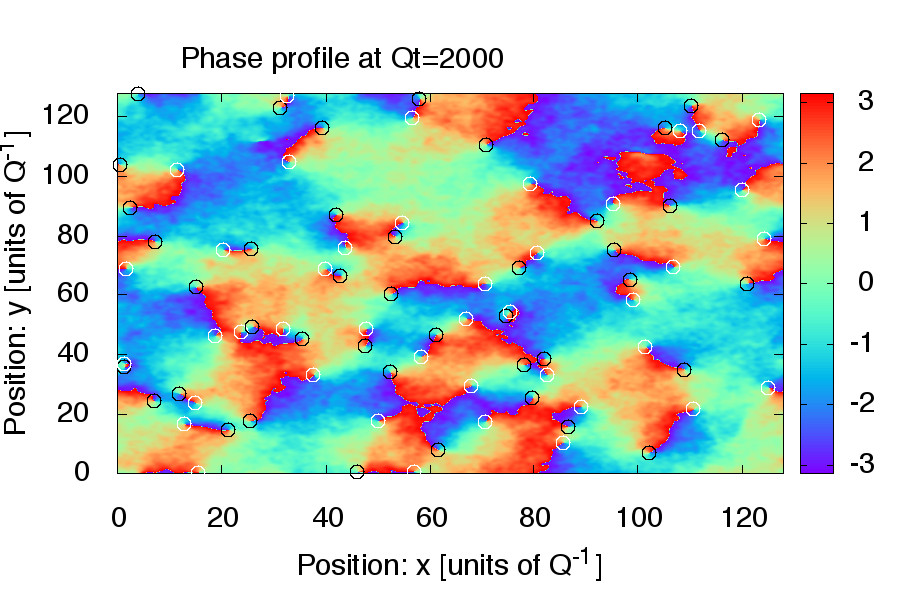}
\includegraphics[width=0.45\textwidth]{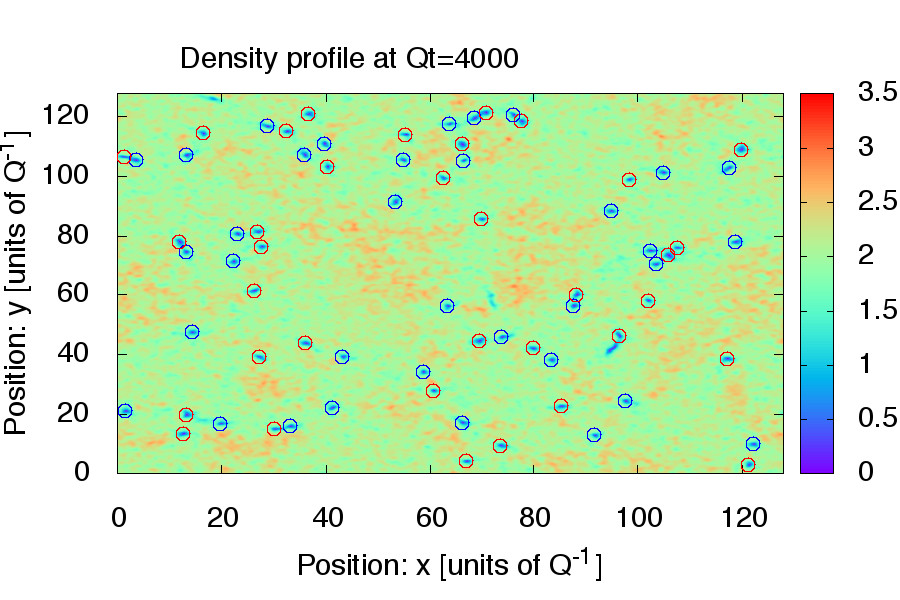}
\includegraphics[width=0.45\textwidth]{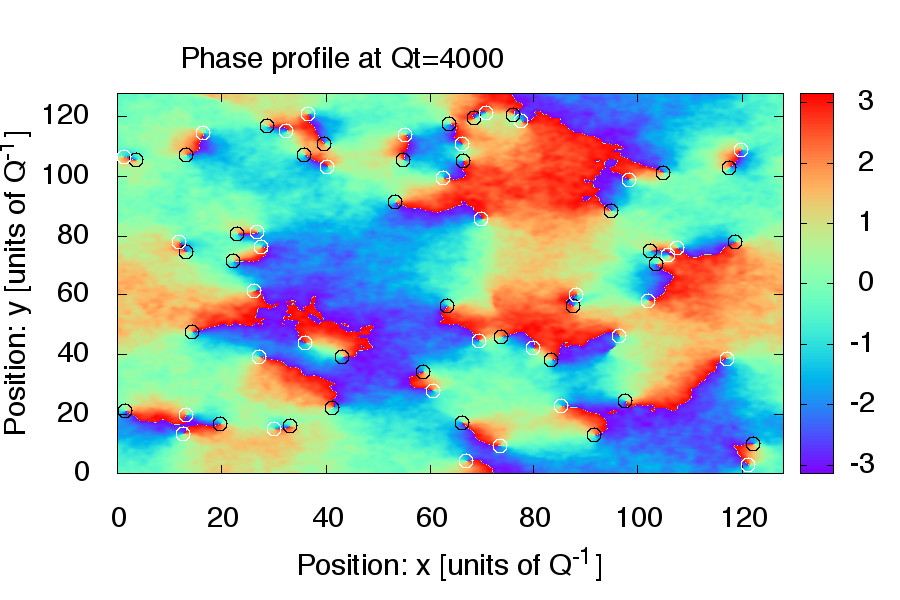}
\caption{Evolution of the profiles of density $|\psi(\xt)|^2$ (left) and the phase angle $\text{arg}(\psi(\xt))$ (right) of the nonrelativistic infrared field $\psi$ obtained by mapping the cooled relativistic field configurations. One clearly observes the formation of long lived vortex and antivortex defects highlighted by oppositely colored circles.}
\label{fig-GPConfigs}       
\end{figure*}

\begin{figure*}[t!]
\centering
\includegraphics[width=\textwidth]{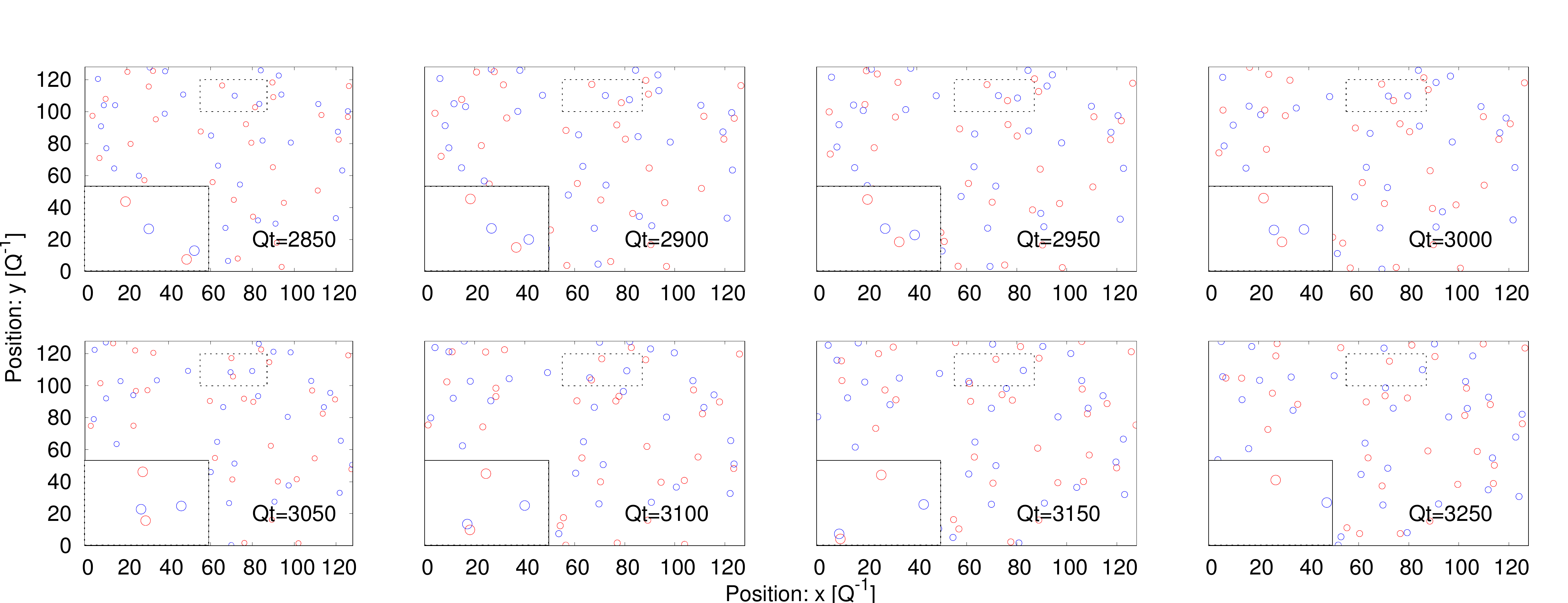}
\caption{Evolution of the distribution of vortices (blue) and anti-vortices (red) at intermediate simulation times. Highlighted in the inset (lower left corner) is a zoom of the dynamics of a vortex-antivortex annihilation process taking place in the dashed rectangular region.}
\label{fig-VortexAnnihilation}       
\end{figure*}

\subsection{Defect structure of nonrelativistic field modes}

Previous studies of the dynamics of the GP equation in two dimensions have demonstrated the importance of vortex defects associated with the global $U(1)$ symmetry of the GP equation (see  \cite{Karl:2016wko,Schole:2012kt,Nowak:2011sk}). We may therefore expect to observe similar features for the nonrelativistic infrared excitations of the single component scalar theory, which in accordance with the discussion in Sec.~\ref{sec:mgp} also feature an approximate $U(1)$ symmetry \cite{Moore:2015adu}. In order to establish a meaningful comparison with the literature, we can express our results in terms of the vortex healing length, defined as
\begin{eqnarray}
\xi_{h} =\left(2 m g n_{\psi}\right)^{-1/2}\;,
\end{eqnarray}
where $g=\lambda/(8m^2)$ is the Gross-Pitaevski coupling. When expressed in terms of the simulation parameters for the relativistic field theory, the healing length is given by
\begin{eqnarray}
Q\xi_{h}=2 \sqrt{\frac{m}{Q}} \left[   \frac{(\lambda/Q) n_{\psi}}{Q^2} \right]^{-1/2}\,,
\end{eqnarray}
and the typical values encountered in our simulations are compactly summarized in Table~\ref{tab-hl}. 

Some examples of the profiles of the field $\psi$ -- obtained from the relativistic scalar fields according to Eqs.~(\ref{eq:psi0}) and (\ref{eq:psi1}) -- are displayed in Fig.~\ref{fig-GPConfigs}, where we present profiles of the magnitude $|\psi|$ and phase-distribution $\mathrm{arg}(\psi)$ at different stages of the nonequilibrium evolution. We find that at all times of the simulation, the field $\psi$ clearly exhibits vortex defects characterized by a sharp minimum of the density with an approximately uniform phase wrap. In order to efficiently detect such vortex defects for a given field configuration, we have developed the following vortex finding algorithm:
we first check for the positions $\xt_{C}$ where the modulus of the field $|\psi(\xt_{C})|$ is smaller than a certain threshold value, $|\psi(\xt_{C})| < 0.1 \langle |\psi(\xt_{C})| \rangle_{\mathrm{V}}$, with the average taken over the system volume. These positions then serve as candidates for vortex cores. Subsequently, we calculate the phase-winding $\mathcal{C}$ around the candidate vortex core, by calculating the line integral of the gradient of the phase $\arg(\psi(\xt))$ of the vortex core. In order to avoid problems with the branch-cut of the $\arg$ function, the line integral $\mathcal{C}$ is calculated as $\mathcal{C}=\sum_{i} \sin\Big[\arg(\psi(\xt_{i+1}))-\arg(\psi(\xt_{i}))\Big]$ along a sequence of points $\xt_i=\xt_{C}+d(\cos(\phi_i),\sin(\phi_i))$ ordered by coordinate space angles $\phi_{i+1} > \phi_{i}$ with a distance $d=|\xt_{C}-\xt_{i}|=1/Q$ around the vortex core. We find that in most cases the vortex charge $\mathcal{C}$ obtained in this way is very close to $\pm2\pi$, typically to percent level accuracy. However to avoid possible misidentification, which arise for example due to regions in space where the density is small but there is no vortex, we apply a quality cut $|\mathcal{C}\pm 2\pi| < \pi/4$ for the identification of the vortex/antivortex. Ultimately, to avoid double counting, we also remove duplicate entries based on the requirement that individual vortices with the same charge at positions $\xt_{C}^{1}$ and $\xt_{C}^{2}$  should be separated by at least $|\xt_{C}^{1}-\xt_{C}^{2}|>1/Q$. 

Vortex defects identified by our algorithm are highlighted by circles in Fig.~\ref{fig-GPConfigs}, demonstrating our ability to efficiently track the defects. By analyzing the vortex dynamics in more detail, we find that many characteristic features previously reported in simulations of the two-dimensional Gross Pitaevskii equation \cite{Karl:2016wko,Schole:2012kt,Nowak:2011sk}, also emerge in our simulations. We find for example that after a rapid initial stage, where vortex-antivortex annihilation occurs frequently, the dynamics slows down significantly towards later times, as the defects tend to organize in vortex-anti vortex pairs. In this regime, vortex-antivortex annihilation occurs predominantly via the interaction of multiple vortices, such as the interaction of a vortex-antivortex pair with an ``unbound'' vortex, or via the interaction of two vortex-antivortex pairs. One example of such a process is depicted in Fig.~\ref{fig-VortexAnnihilation}, where we show the distribution of vortices and anti-vortices at different times $Qt=2850 -3250$. In the inset, we present a zoom of the dynamics around the point $(x,y)\sim(80,110)$, where an annihilation process takes place following the interaction of two vortex-antivortex pairs.

\begin{figure}[t!]
\centering
\includegraphics[width=\textwidth]{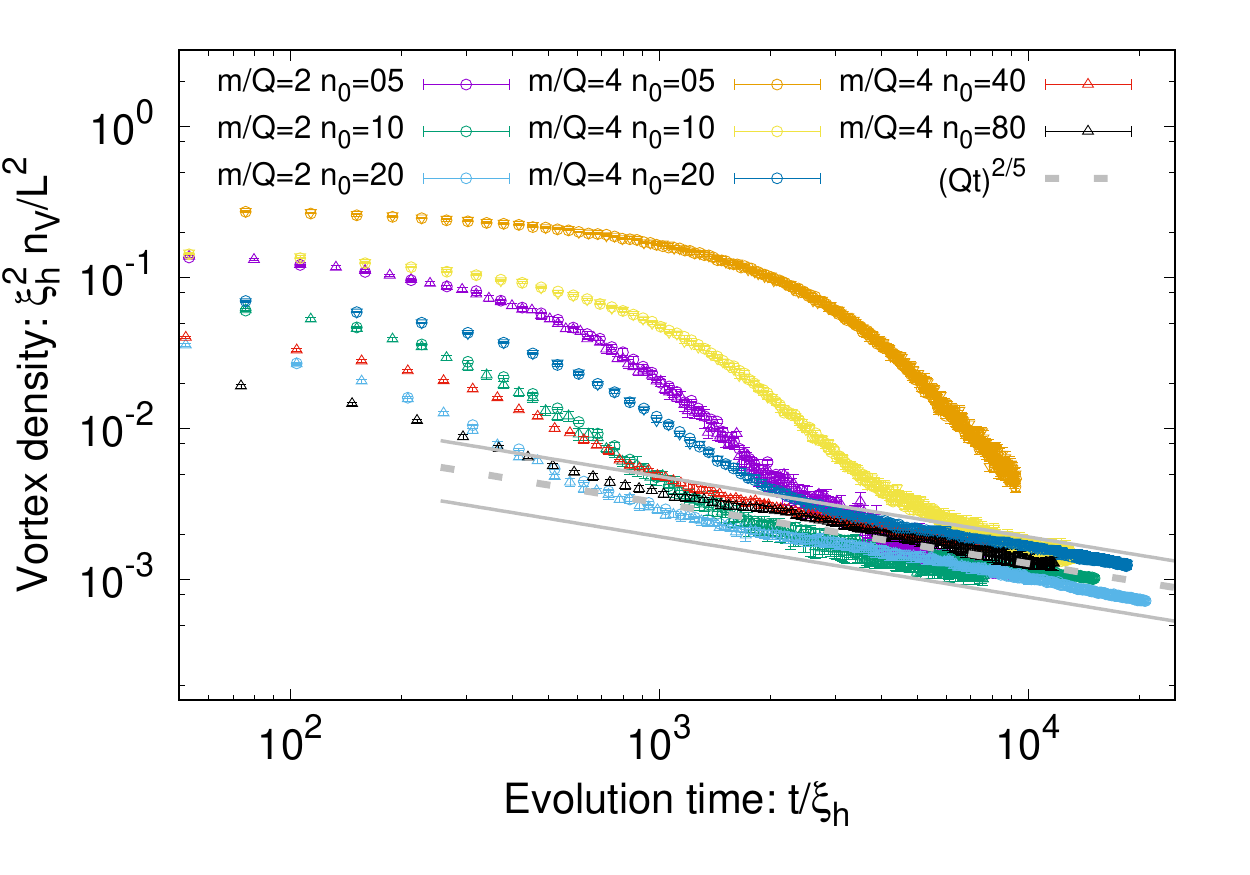}
\includegraphics[width=\textwidth]{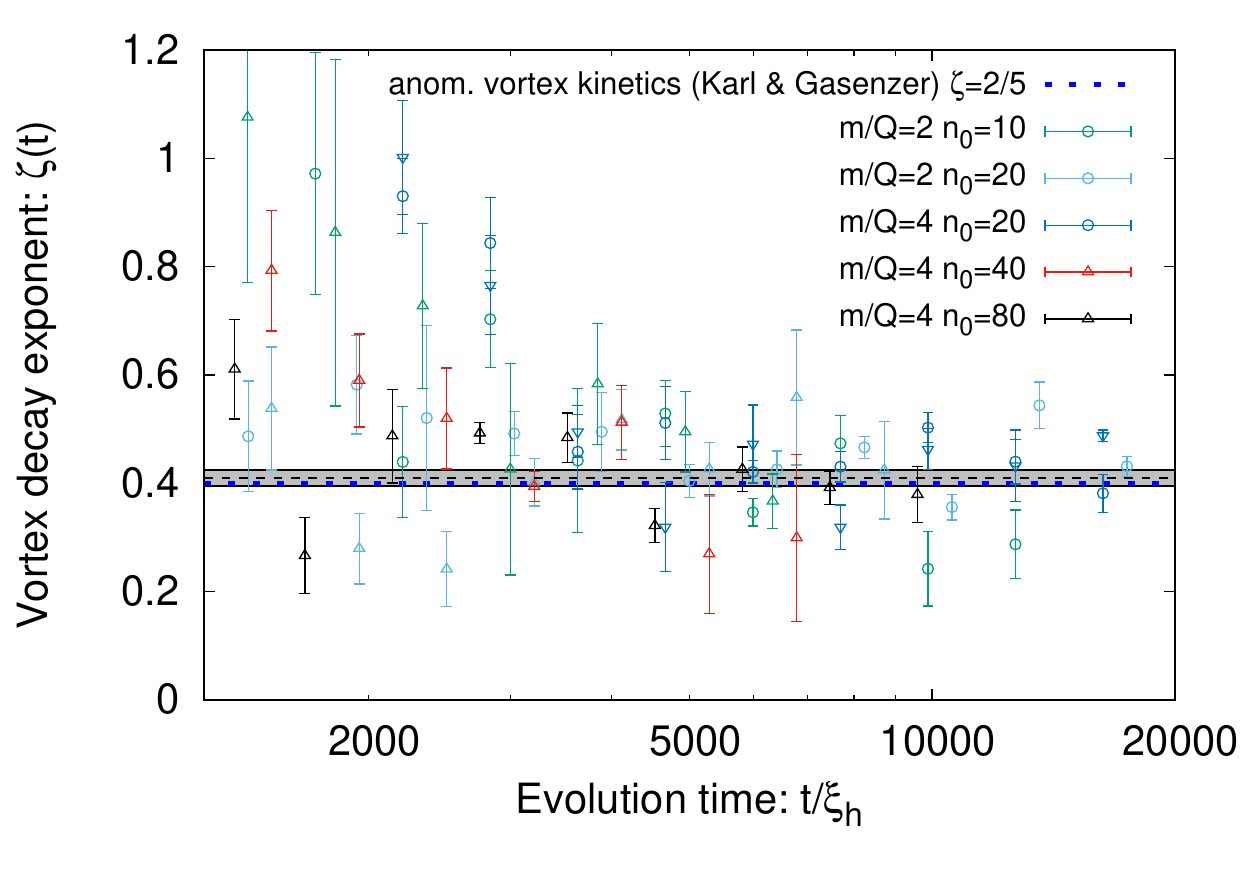}
\caption{(Top) Evolution of the vortex density $\xi_{h}^2 n_{V}/L^2$ as functions of time $t$ in the unit of the healing length. (Bottom) Extraction of the scaling exponent $\zeta$ characterizing the power law decay $n_{V} \sim t^{-\zeta}$ observed at late times.}
\label{fig-VortexDensity}       
\end{figure}

\subsection{Statistical properties of vortex dynamics}
Beyond the investigation of the defect structure of individual configurations, it is also useful to analyze the statistical properties of the defect dynamics. Our result for the average vortex density $\xi_{h}^2 n_{V}/L^2$ (expressed in the unit of the inverse healing length squared) is presented in Fig.~\ref{fig-VortexDensity}, where we show results for different values of $m/Q$ and the initial overoccupancy parameter $n_0$ as a function of the evolution time $t/\xi_{h}$. We find that following a transient regime which lasts between $10^{3} \xi_{h}$ to $10^{4} \xi_{h}$ (depending on $n_0$ and $m/Q$), the vortex density approaches a power law decay form $n_{V} \sim t^{-\zeta}$ with $\zeta\approx0.4$ indicated by the gray band. In order to extract a more precise estimate of the power law index $\zeta$, which governs the decay of the vortex density, we extract the logarithmic derivative
\begin{eqnarray}
\zeta(t)=-\frac{d\log(n_{V})}{d\log(t)}\,,
\end{eqnarray}
shown in the right panel of Fig.~\ref{fig-VortexDensity}. Performing a simultaneous fit to all data points with $t/\xi_{h} >5000$ shown in the right panel, we estimate
\begin{eqnarray}
\label{eq:zetaval}
\zeta=0.41\pm0.015\,.
\end{eqnarray}
Our result in Eq.~(\ref{eq:zetaval}) is in excellent agreement with the values of $\zeta$ reported in \cite{Karl:2016wko} for the vortex dynamics in a two-dimensional superfluid near the strongly anomalous nonthermal fixed point, as is indicated by the dashed blue line in Fig.~\ref{fig-VortexDensity}.

\begin{figure*}[t!]
\centering
\includegraphics[width=0.45\textwidth]{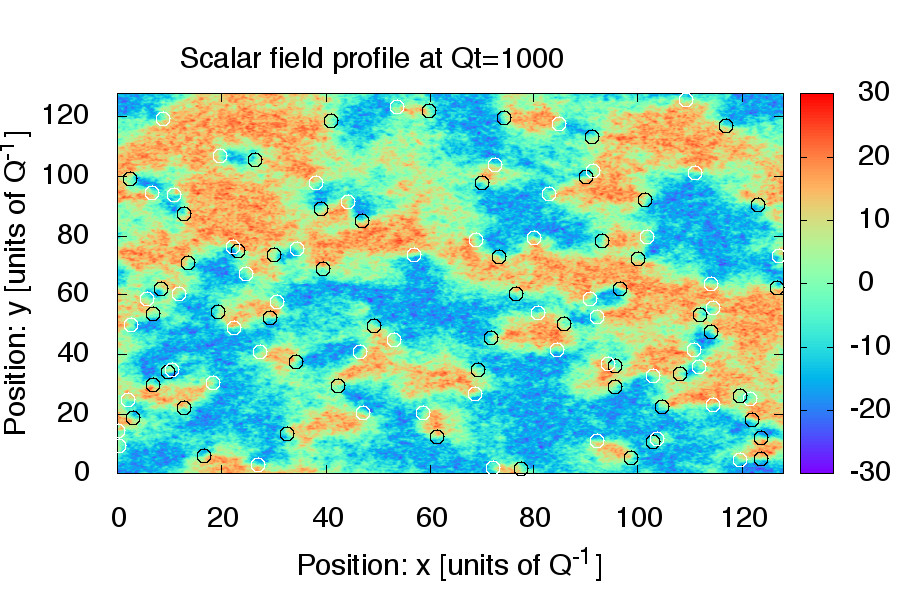}
\includegraphics[width=0.45\textwidth]{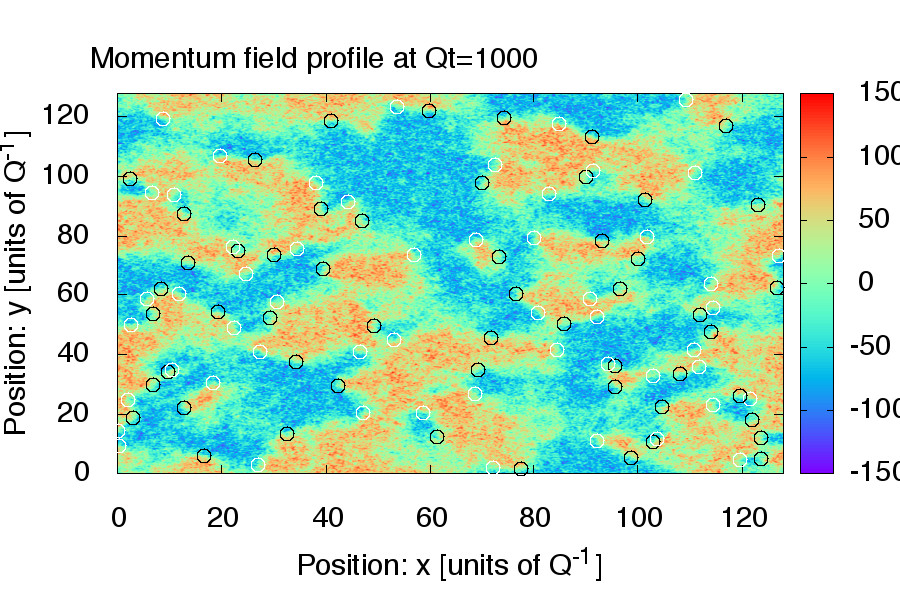}
\includegraphics[width=0.45\textwidth]{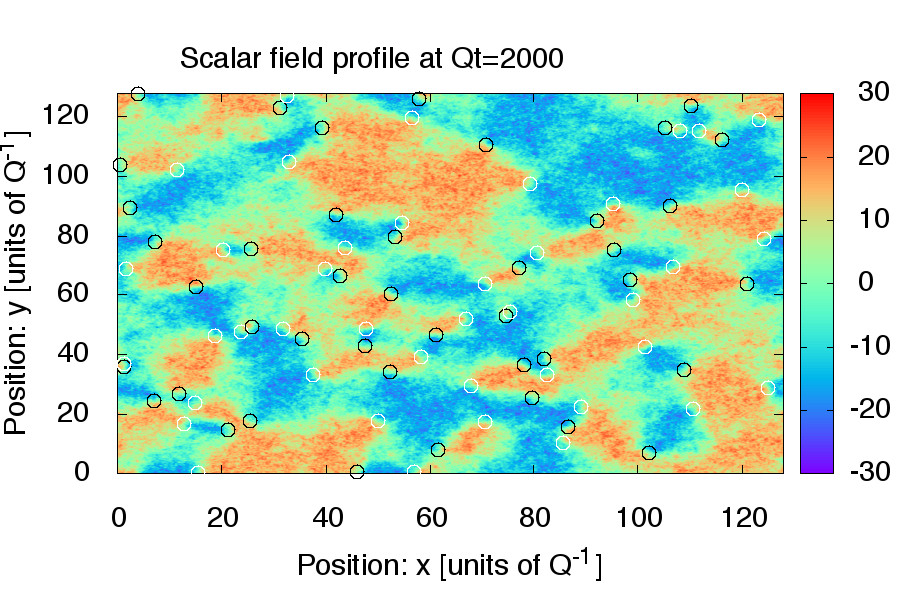}
\includegraphics[width=0.45\textwidth]{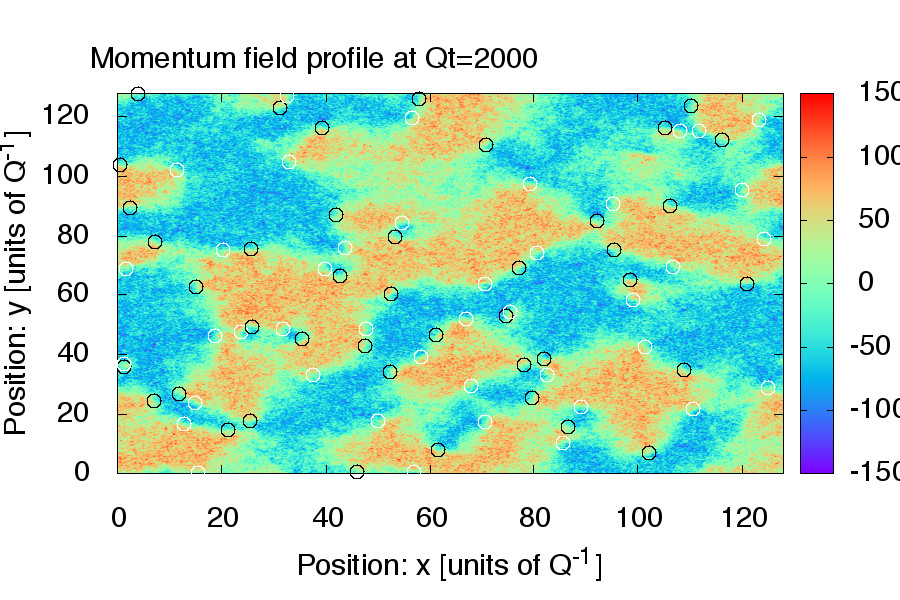}
\includegraphics[width=0.45\textwidth]{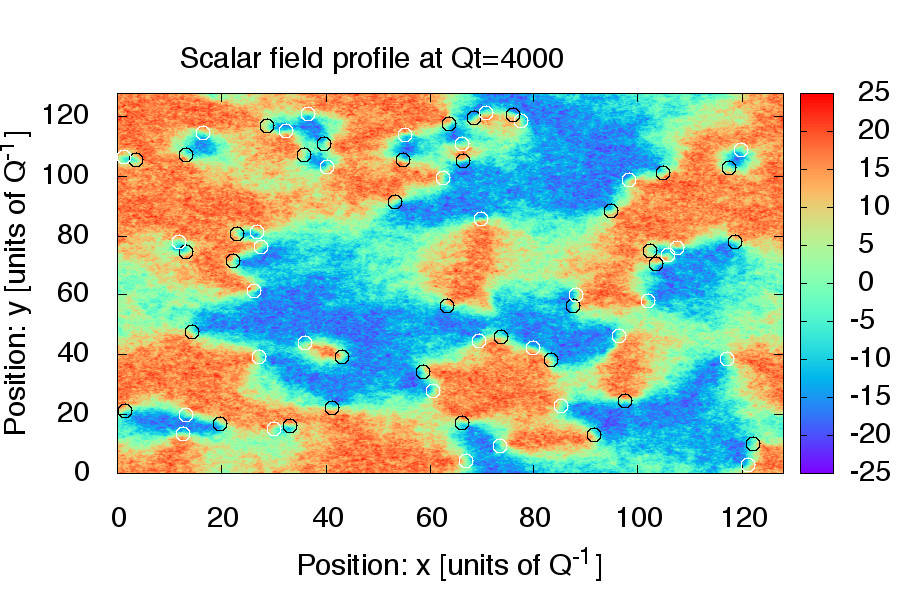}
\includegraphics[width=0.45\textwidth]{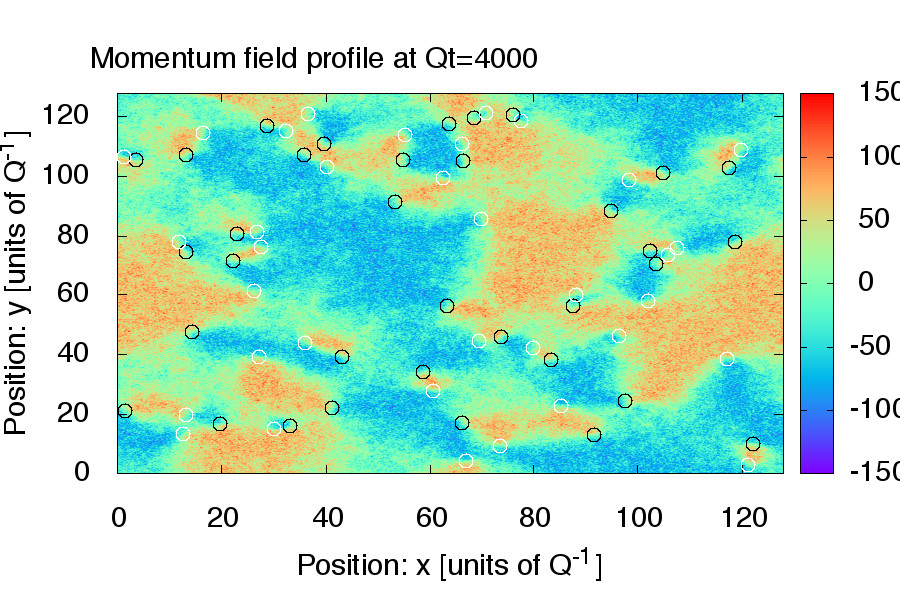}
\caption{Evolution of the profiles of relativistic field configurations for the scalar field $\phi$ (left) and its conjugate momentum field $\pi$ (right). Domain structures of both fields are clearly visible. Intersections of domain walls of $\phi$ and $\pi$ correspond to location of vortex defects.}
\label{fig-RelConfigs}       
\end{figure*}

\subsection{Defect structure of original relativistic field}
It is also interesting to investigate how features of the defect structure manifest themselves at the level of the original fields $\phi$ and $\pi$. From the spatial profile plots of the relativistic fields presented in Fig.~\ref{fig-RelConfigs}, one observes the expected emergence of a $Z_{2}$ domain structure for both $\phi$ and $\pi$, with the domain size $\ell_{\mathrm{domain}}$ growing as a function of time. We find that the position of vortex defects discussed previously in terms of the modulus and phase of the nonrelativistic $\psi$ field corresponds to intersection points of the domain walls of the original $\phi$ and $\pi$ fields. Hence the growth of the characteristic domain size $\ell_{\mathrm{domain}}$ is limited by the presence of vortex defects and can be expected to follow the growth of the average distance between vortices given by
\begin{eqnarray}
\label{eq:domaingrowth}
\ell_{\mathrm{domain}}\sim \ell_{V} \sim n_{V}^{-1/2} \sim t^{\zeta/2}\, .
\end{eqnarray}
This defect picture explains some of the features observed in the infrared sector of the relativistic scalar theory in terms of the underlying dynamics of the nonrelativistic degrees of freedom. Since the single particle spectrum discussed in Sec.~\ref{sec:spectra-ntfp}, is essentially a two-point correlation function of the $\phi$ and $\pi$ fields, we can relate the characteristic momentum scale $p_{\mathrm{IR}}$  of the spectrum, marking the transition from the $p^{-4}$ power law to the $~constant$ behavior in the deep infrared, with the inverse of the typical domain size $p_{\mathrm{IR}}\sim\ell^{-1}_{\mathrm{domain}}$. The $p^{-4}$ power law in the spectrum should then be  attributed to the strongly correlated dynamics within a domain whereas the $constant$ behavior in the deep infrared can be attributed to a statistical average over uncorrelated domains. See also Ref. \cite{Moore:2015adu} for a related discussion. 

Since the scaling exponents for the regime with $p> p_{\mathrm{IR}}$ and $p<p_{\mathrm{IR}}$ should match if the dynamics is self-similar throughout, we obtain $\zeta=2\beta$.  This expectation is clearly corroborated by our numerical results in Eqs.~(\ref{eq:betaval}) and (\ref{eq:zetaval}). Hence the infrared dynamics of the relativistic  scalar field theory can be understood efficiently in terms of the dynamics of the vortex defects of the nonrelativistic infrared fields.

\begin{figure}[t!]
\centering
\includegraphics[width=\textwidth]{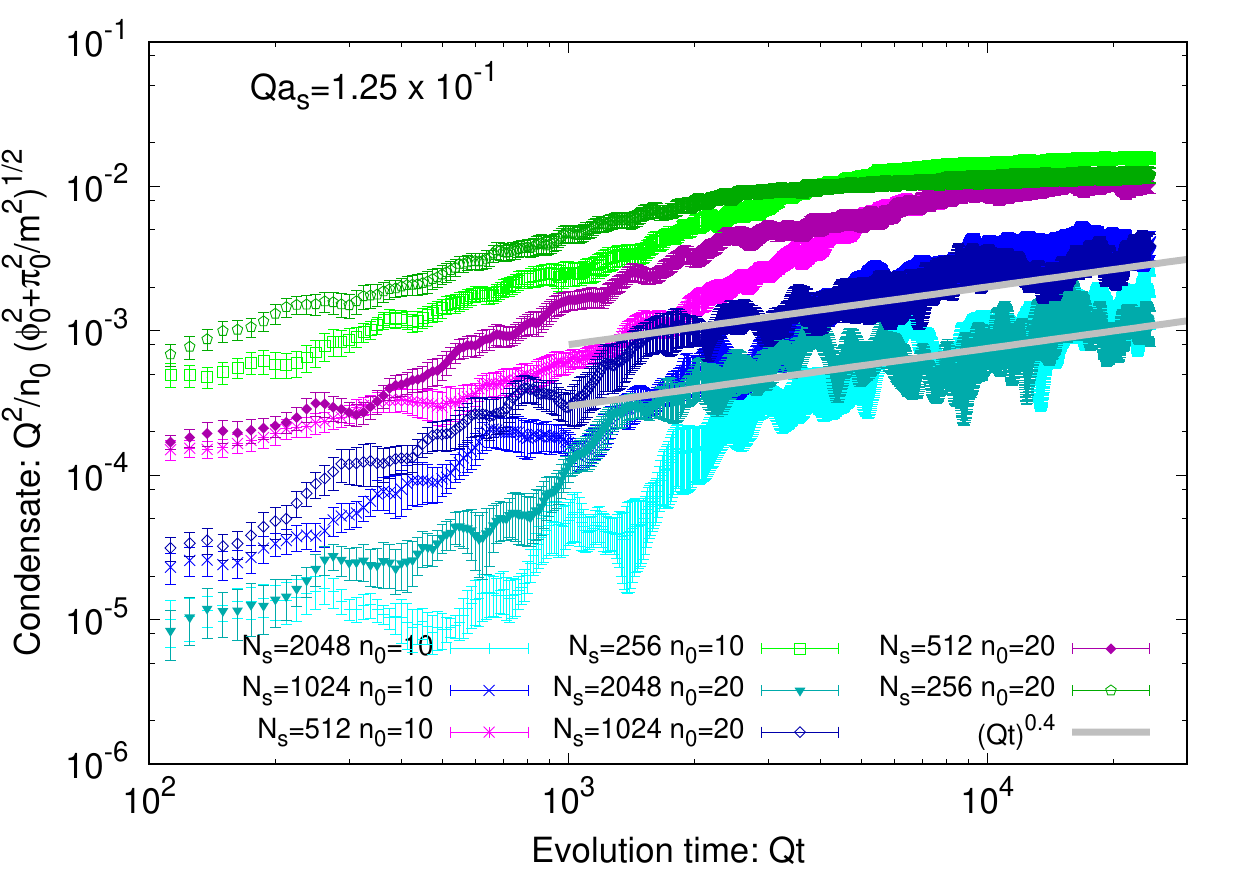}
\includegraphics[width=\textwidth]{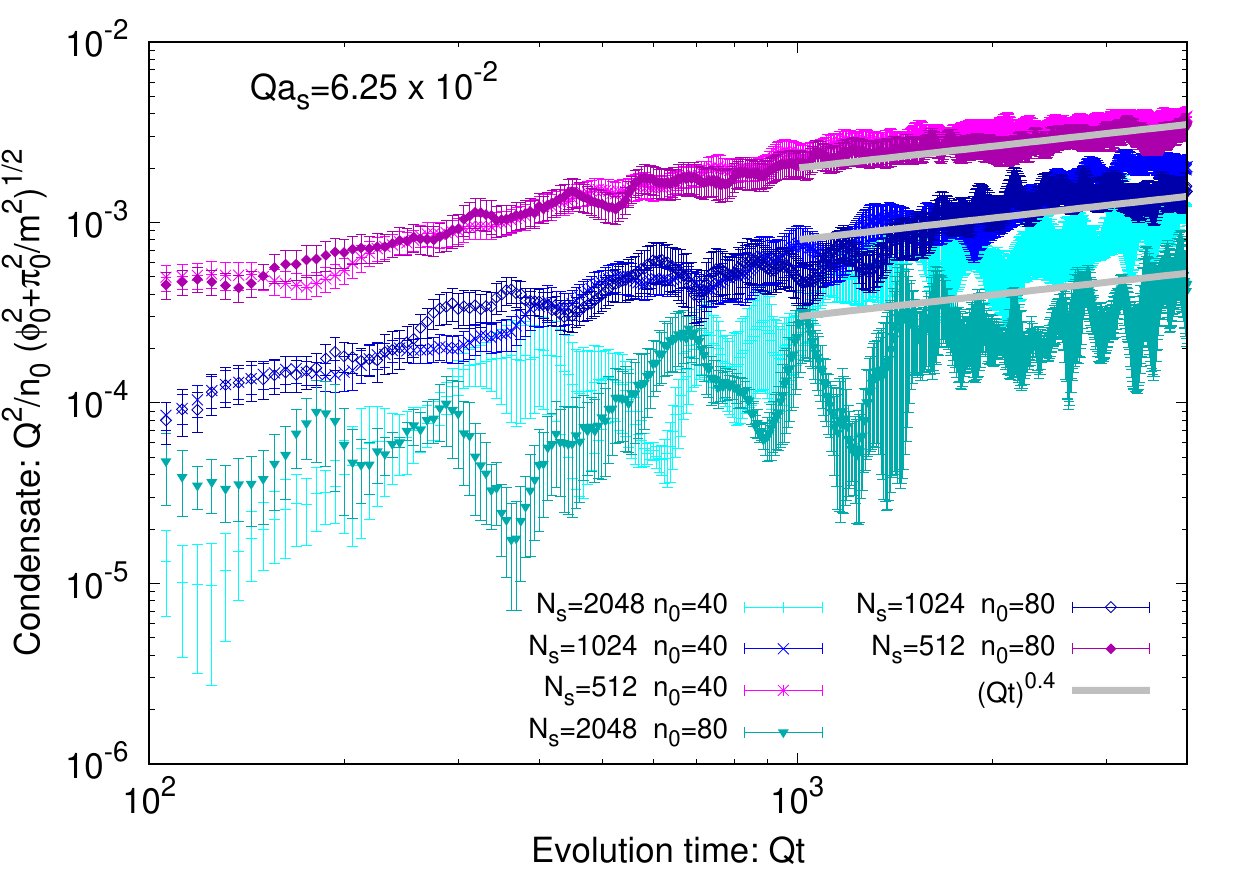}
\caption{Evolution of the condensate observables $\tilde{\phi}^2(p=0) + \tilde{\pi}^2(p=0)/m^2$ for $m/Q=4$ for different physical volumes from $256^2$ to $2048^2$ and varying amount of initial overoccupancy $n_0=10-80$.}
\label{fig-RelConfigs-1}       
\end{figure}

One may also deduce from the structure of the field configurations in Fig.~\ref{fig-RelConfigs-1} that the presence of such vortex defects prohibits the formation of a Bose condensate corresponding to a single coherent domain across the entire  volume of the two-dimensional system. We can confirm this expectation by investigating the dynamics of the zero mode $\tilde{\phi}^2(p=0) + \tilde{\pi}^2(p=0)/m^2$, for different physical volumes and for various choices of simulation parameters in Fig.~\ref{fig-RelConfigs-1}. 
Even though we are statistics limited, we observe that a strong volume dependence of the correlation function persists over the course of the entire simulation -- this indicates the absence of a Bose condensate \cite{Berges:2012us}.  From the arguments of \cite{Orioli:2015dxa}, one may then expect that  -- for sufficiently large volumes -- the growth of the condensate is described by a power law $t^{\alpha}$ such that the condensation time diverges as $t^{1/\alpha}$. One can also arrive at the same conclusion by considering the phase-ordering kinetics described above. Albeit, as noted, the current statistics have significant uncertainties, we see from  Fig.~\ref{fig-RelConfigs-1} that the grey bands with a power law dependence of $t^{0.4}$ are consistent with the numerical data. Since $\alpha=2\beta$ with $\beta = 0.24 \pm 0.08$, our numerical results are in line with the expectation of \cite{Berges:2012us} for the power law growth of the Bose condensate.

\section{Summary and outlook}

In this paper, we demonstrated a formal map between the infrared structure of an $N=1$ relativistic self-interacting scalar field theory and 
the Gross-Pitaevskii (GP) theory for nonrelativistic fields, which is widely employed as a model theory describing the behavior of superfluids. This map follows from observing that any real scalar field can be Fourier transformed to momentum space, where it can be decomposed into a complex valued creation/annihilation variables $a_{\mathbf{p}}$  and $a_{\mathbf{p}}^*$. By using alternative canonical variables $b_{\mathbf{p}}$ and $b_{\mathbf{p}}^*$, which dress the original $a_{\mathbf{p}}$ and $a_{\mathbf{p}}^*$ functions, one can transform out nonresonant quartic terms to obtain a Hamiltonian
with only quartic resonant interactions. The resonant term has an equal number of $b_{\mathbf{p}}$ and $b_{\mathbf{p}}^{*}$ terms and therefore conserves particle number that is manifest in the global U(1) symmetry of the Hamiltonian in the dressed  $b_{\mathbf{p}}$ and $b_{\mathbf{p}}^*$ canonical variables. This Hamiltonian is the GP Hamiltonian for the nonrelativistic complex field $\psi(\mathbf{x})$, which is the  the Fourier transform of $b_{\mathbf{p}}$.  Higher order interaction terms with more than four $b_{\mathbf{p}}$ or $b_{\mathbf{p}}^{*}$ are generated by this mapping. However these terms are suppressed by powers of $\lambda \psi^{2}(\mathbf{x})/m^{3}$ or $\lambda \phi ^2/m^2$ in  the nonrelativistic limit or large mass limit. In addition to computing higher order corrections to the GP Hamiltonian, this map allows us to include within the GP framework the role of fluctuations that can be treated systematically in the underlying relativistic field theory. Perhaps most importantly, it also allows us to dynamically study the coupling of superfluid modes to normal ultraviolet modes out of equilibrium. 

We reported on numerical simulations of the far out of equilibrium temporal evolution of overoccupied fields in the $N=1$ scalar field theory. We observed that self-similar behavior characteristic of nonthermal fixed points develops as a function of time. This is characterized by different spectral indices for the quasi-stationary distributions in different inertial ranges, both in the UV and IR sectors of the theory. We focused on the dynamics in the IR sector and were able to do this by systematically removing UV modes using a well developed lattice cooling technique. Exploiting the classical map we established between the single component relativistic scalar field $\phi$ and the Gross-Pitaevski field $\psi$, we further analyzed the large scale structure of the effective nonrelativistic field configurations. 

The complex scalar fields $\psi$ we extracted have a spatial structure that includes the presence of vortex-antivortex pairs. At early times and high occupancies, they densely occupy the system. However subsequently the scattering and annihilation of these vortices leads to a dilution of the vortex density with a characteristic power law decay in time. This coarsening dynamics is observed in the spatial structure of the relativistic scalar field and its conjugate momentum field where the vortex defects correspond to intersections of the domain walls separating distinct domains of phase coherent fields. The characteristic scale associated with the inverse domain size $p_{\mathrm{IR}}\sim\ell^{-1}_{\mathrm{domain}}$ separates spectral properties of distributions in the deep infrared from those at momenta larger than this scale. In the former case, the incoherent averaging over multiple domains gives rise to a spectral index that is approximately vanishing. Conversely, for momenta $p>p_{\mathrm{IR}}$ the spectrum is determined by the dynamics within a single domain that generates a $1/p^4$ behavior. 

Within the accuracy of our simulations, our observations are consistent with numerical simulations of the Gross-Pitaveski equation by Karl and Gasenzer.  By studying the dynamics of a quenched two-dimensional Bose gas, they observed a NTFP with ``strongly anomalous'' scaling properties that are distinct from the predictions of a vertex resummed kinetic theory. While the phase kinetic evolution associated with this nonthermal fixed point is distinct from the near-equilibrium phase ordering kinetics~\cite{Bray:1994zz}, it was pointed out by Karl and Gasenzer that the anomalous scaling properties can be understood in terms of a modified phase kinetic picture of correlated vortex defects. From our analysis of the effective nonrelativistic infrared degrees of freedom, we conclude that the same dynamics also takes place in our simulations of the two-dimensional massive relativistic scalar theory. This suggests that (hidden) topological defects can play an important role for understanding the properties of NTFPs.

While we have only presented numerical results for the single component $(N=1)$ theory in (2+1)-dimensions, our relativistic field theory framework is more general and can be exploited to explore in greater detail the effective degrees of freedom governing the out-of-equilibrium dynamics of Bose superfluids. Though more numerically cumbersome, our techniques can be extended in a straightforward way to (3+1)-dimensions, where the defect structure is likely to be distinct from that in (2+1)-dimensions \cite{Moore:2015adu}. One of the more important questions is to clarify the relative importance of quasiparticle excitations and topological defects in determining the nonequilibrium scaling properties for different number of field components ($N=1,2,\cdots$) and the dimensionality of the system. 

Interestingly, the methods we have applied here can potentially be applied to understand the very high vorticity of the strongly interacting matter produced in heavy-ion collisions at the Relativistic Heavy-Ion Collider (RHIC)~\cite{Abelev:2007zk,STAR:2017ckg}. At the energies where the high vorticity is observed through the decays of 
polarized hyperons ~\cite{Becattini:2016gvu,Fang:2016vpj,Li:2017slc,Wang:2017jpl,Siddique:2017ddr}, the  matter is a baryon rich fluid whose late time evolution is described by hydrodynamics. How the global vorticity induced at infrared scales is dynamically transferred to the ultraviolet scale of the viscosity in the hydrodynamical fluid on the rapid time scales of the collision is not well understood. The ideas developed here provide a possible explanation. Firstly, because of the relativistic nature of the collision, there is a natural separation between the longitudinal and transverse dynamics in the fluid so key aspects of the dynamics take place in the two transverse dimensions. Secondly, the relativistic scalar theory we have considered here is a good model for a baryon rich fluid~\cite{Berges:1998rc,Alford:2012vn,Son:2004iv} and out-of-equilibrium phenomena in the vicinity of a critical end point in the QCD phase diagram have been explored in such scalar field theories~\cite{Berdnikov:1999ph,Berges:2009jz,Mukherjee:2015swa}; indeed, self-similar behavior analogous to that discussed here is observed in such studies~\cite{Mukherjee:2016kyu}. The turbulent processes we have discussed here are efficient mechanisms to transfer the global angular momentum induced in a noncentral heavy-ion collision to the ultraviolet scale described by nearly ideal hydrodynamics. It would be particularly interesting in this context to understand the possible role of vortex structures in generating the polarization of particles on the microscopic scales of the fluid. These exciting possibilities will be pursued in forthcoming work.

\textit{Acknowledgments}. 
We would like to thank Juergen Berges, Kirill Boguslavski, Thomas Gasenzer, Mohammad Hossein Namjoo
and Hong-zhong Wu for valuable comments and discussions. 
J.D. and Q.W. are supported in part by the Major State Basic Research Development Program (973 Program) 
in China under Grant No. 2014CB845400. 
Q.W. is supported in part by the Major State Basic Research Development Program (973 program) 
in China under the Grant No. 2015CB856902 and by the National Natural 
Science Foundation of China (NSFC) under the Grant No. 11535012. 
Q.W. was supported jointly by China Scholarship Council and the nuclear theory group
of Brookhaven National Laboratory as a senior research fellow when
this work was initiated and a large part of it was completed. 
This work is part of and supported by the DFG Collaborative Research Centre 
``SFB 1225 (ISOQUANT)''. 
R.~V.'s and S.S's research is supported by the U.\ S.\ Department of Energy Office of Science, 
Office of Nuclear Physics, under contracts No.\ DE-SC0012704 (R.V.) and DE-FG02-97ER41014 (S.S.). 
R.V's work is supported within the framework of the Beam Energy Scan Theory (BEST) Topical Collaboration. 
He would also like to thank ITP Heidelberg and the Alexander Von Humboldt Foundation 
for support and ITP Heidelberg for their kind hospitality. 

\appendix

\section{Some properties of canonical transformation}
\label{sec:canonical}
In this appendix, we shall derive some properties
of canonical transformation that are used in Section \ref{sec:can-var}. 

We will first show that the Poisson brackets are invariant under a change of
canonical basis. Towards this end, we will show that one can rewrite the Poisson brackets
in Eq.~(\ref{eq:poisson-a}) in a new canonical basis $b_{\mathbf{p}}$,
\begin{eqnarray}
&&\left\{ F,G\right\} _{a}  =  \int[d^{3}\mathbf{k}]\left[\frac{\partial F}{\partial a_{\mathbf{k}}}\frac{\partial G}{\partial a_{\mathbf{k}}^{*}}-\frac{\partial F}{\partial a_{\mathbf{k}}^{*}}\frac{\partial G}{\partial a_{\mathbf{k}}}\right]\nonumber \\
 & = & \int[d^{3}\mathbf{k}][d^{3}\mathbf{k}_{1}][d^{3}\mathbf{k}_{2}]\nonumber \\
 &  & \left\{ \left[\frac{\partial b_{\mathbf{k}1}}{\partial a_{\mathbf{k}}}\frac{\partial F}{\partial b_{\mathbf{k}1}}+\frac{\partial b_{\mathbf{k}1}^{*}}{\partial a_{\mathbf{k}}}\frac{\partial F}{\partial b_{\mathbf{k}1}^{*}}\right]\left[\frac{\partial b_{\mathbf{k}2}}{\partial a_{\mathbf{k}}^{*}}\frac{\partial G}{\partial b_{\mathbf{k}2}}+\frac{\partial b_{\mathbf{k}2}^{*}}{\partial a_{\mathbf{k}}^{*}}\frac{\partial G}{\partial b_{\mathbf{k}2}^{*}}\right]\right.\nonumber \\
  &  & -\left.\left[\frac{\partial b_{\mathbf{k}1}}{\partial a_{\mathbf{k}}^{*}}\frac{\partial F}{\partial b_{\mathbf{k}1}}+\frac{\partial b_{\mathbf{k}1}^{*}}{\partial a_{\mathbf{k}}^{*}}\frac{\partial F}{\partial b_{\mathbf{k}1}^{*}}\right]\left[\frac{\partial b_{\mathbf{k}2}}{\partial a_{\mathbf{k}}}\frac{\partial G}{\partial b_{\mathbf{k}2}}+\frac{\partial b_{\mathbf{k}2}^{*}}{\partial a_{\mathbf{k}}}\frac{\partial G}{\partial b_{\mathbf{k}2}^{*}}\right]\right\} \nonumber \\
 & = & \int[d^{3}\mathbf{k}_{1}][d^{3}\mathbf{k}_{2}]\nonumber \\
 &  & \left[\left\{ b_{\mathbf{k}1},b_{\mathbf{k}2}\right\} _{a}\frac{\partial F}{\partial b_{\mathbf{k}1}}\frac{\partial G}{\partial b_{\mathbf{k}2}}+\left\{ b_{\mathbf{k}1},b_{\mathbf{k}2}^{*}\right\} _{a}\frac{\partial F}{\partial b_{\mathbf{k}1}}\frac{\partial G}{\partial b_{\mathbf{k}2}^{*}}\right.\nonumber \\
 &  & +\left.\left\{ b_{\mathbf{k}1}^{*},b_{\mathbf{k}2}\right\} _{a}\frac{\partial F}{\partial b_{\mathbf{k}1}^{*}}\frac{\partial G}{\partial b_{\mathbf{k}2}}+\left\{ b_{\mathbf{k}1}^{*},b_{\mathbf{k}2}^{*}\right\} _{a}\frac{\partial F}{\partial b_{\mathbf{k}1}^{*}}\frac{\partial G}{\partial b_{\mathbf{k}2}^{*}}\right]\nonumber \\
 & = & \left\{ F,G\right\} _{b}
\end{eqnarray}
where we have required that $b_{\mathbf{k}}$ and $b_{\mathbf{k}}^{*}$ be canonical variables satisfying Eq. (\ref{eq:new-basis}). We list following rules for the Poisson brackets, 
\begin{eqnarray}
\{AB,C\} & = & A\{B,C\}+B\{A,C\},\nonumber \\
\{C,AB\} & = & -\{AB,C\}=A\{C,B\}+B\{C,A\},\nonumber \\
\{AB,CD\} & = & BD\{A,C\}+BC\{A,D\} \nonumber \\
&& +AC\{B,D\}+AD\{B,C\},
\label{eq:rule-poisson}
\end{eqnarray}
where $A$, $B$ and $C$ are all momentum functions, and we have
suppressed the basis for above Poisson brackets which can be any basis. 


To see that $\tb_{\mathbf{p}}(z)$ and $\tb_{\mathbf{p}}^{*}(z)$ in Eq.
(\ref{eq:taylor-at0}) satisfy the canonical relations in Eq.~(\ref{eq:new-basis})
for any form of $H_{\mathrm{aux}}(\tilde{b},\tilde{b}^*)\equiv H_{\mathrm{aux}}(\tb_{\mathbf{p}}(z),\tb_{\mathbf{p}}^{*}(z))$, 
we will directly evaluate the Poisson brackets $\{\tb_{\mathbf{p}}(z),\tb_{\mathbf{p}1}^{*}(z)\}_{b}$
and $\{\tb_{\mathbf{p}}(z),\tb_{\mathbf{p}1}(z)\}_{b}$ by Taylor expansion
in $z$:
\begin{eqnarray}
f_{1}(z)&\equiv&\{\tb_{\mathbf{p}}(z),\tb_{\mathbf{p}1}^{*}(z)\}_{b}  \nonumber \\
& = & \{b_{\mathbf{p}},b_{\mathbf{p}1}^{*}\}_{b}+\sum_{n=1}^{\infty}\frac{1}{n!}\left.\frac{d^{n}f_{1}(z)}{d^{n}z}\right|_{z=0}z^{n}\nonumber \\
 & = & (2\pi)^{3}\delta(\mathbf{p}-\mathbf{p}_{1}),\nonumber \\
f_{2}(z)&\equiv&\{\tb_{\mathbf{p}}(z),\tb_{\mathbf{p}1}(z)\}_{b} \nonumber \\
& = & \{b_{\mathbf{p}},b_{\mathbf{p}1}\}_{b}+\sum_{n=1}^{\infty}\frac{1}{n!}\left.\frac{d^{n}f_{2}(z)}{d^{n}z}\right|_{z=0}z^{n}\nonumber \\
 & = & 0\, .\label{eq:bbs-bb}
\end{eqnarray}
As is transparent from the r.h.s of these relations, for the canonical relations to hold, all the coefficients of $z^{n}$ must vanish. To see this, we note the following identities [the proof follows Eq. (\ref{dobbdz})],  
\begin{equation}
i\left.\frac{d^{n+1}f_{1,2}(z)}{d^{n+1}z}\right|_{z=0}  =  \left\{ \left.\frac{d^{n}f_{1,2}(z)}{d^{n}z}\right|_{z=0},H_{\mathrm{aux}}\right\} _{b}.
\label{eq:induction-0}
\end{equation}
We can evaluate the first derivative of $f_{1,2}$ as  
\begin{eqnarray}
&&\left.\frac{df_{1}(z)}{dz}\right|_{z=0}  =  i\left\{ b_{\mathbf{p}},\frac{\partial H_{\mathrm{aux}}}{\partial b_{\mathbf{p}1}}\right\} _{b} \nonumber\\
&&+i\left\{ b_{\mathbf{p}1}^{*},\frac{\partial H_{\mathrm{aux}}}{\partial b_{\mathbf{p}}^{*}}\right\} _{b}=0,\nonumber \\
&&\left.\frac{df_{2}(z)}{dz}\right|_{z=0}  =  -i\left\{ b_{\mathbf{p}},\frac{\partial H_{\mathrm{aux}}}{\partial b_{\mathbf{p}1}^{*}}\right\} _{b} \nonumber\\
&&+i\left\{ b_{\mathbf{p}1},\frac{\partial H_{\mathrm{aux}}}{\partial b_{\mathbf{p}}^{*}}\right\} _{b}=0\,, 
\nonumber\\ 
\label{eq:induction}
\end{eqnarray}
where we have implied $H_{\mathrm{aux}}\equiv H_{\mathrm{aux}}(b,b^*)=\left. H_{\mathrm{aux}}(\tilde{b},\tilde{b}^*)\right|_{z=0}$. 
In the r.h.s of two equalities in Eq. (\ref{eq:induction}), 
we have used Eq.~(\ref{dbtdz}). By induction, if we assume the coefficient of $z^{n}$ is vanishing,
we see that the coefficient of $z^{n+1}$ is also vanishing from 
Eq. (\ref{eq:induction-0}). We can therefore conclude that all coefficients
of $z^{n}$ for any $n>0$ are vanishing, completing the proof of Eq.~(\ref{eq:bbs-bb}). 

We shall now provide a proof of Eq. (\ref{eq:h-new}). Define a 
quantity $O(\tb_{\mathbf{p}}(z),\tb_{\mathbf{p}}^{*}(z))$, a
functional of $\tb_{\mathbf{p}}(z)$ and $\tb_{\mathbf{p}}^{*}(z)$ in
Eq. (\ref{eq:taylor-at0}), whose virtual time evolution is governed
by $H_{\mathrm{aux}}$. We then have 
\begin{eqnarray}
\left.\frac{dO(\tilde{b},\tilde{b}^{*})}{dz}\right|_{z=0} & = & \int[d^{3}\mathbf{k}]\left[\frac{\partial O(\tilde{b},\tilde{b}^{*})}{\partial\tilde{b}_{\mathbf{k}}}\frac{d\tilde{b}_{\mathbf{k}}}{dz}+\frac{\partial O}{\partial b_{\mathbf{k}}^{*}}\frac{db_{\mathbf{k}}^{*}}{dz}\right]_{z=0}\nonumber \\
 & = & -i\int[d^{3}\mathbf{k}]\left[\frac{\partial O(b,b^{*})}{\partial b_{\mathbf{k}}}\{b_{\mathbf{k}},H_{\mathrm{aux}}\}_{b} \right. \nonumber \\
 && \qquad \qquad +\left.\frac{\partial O(b,b^{*})}{\partial b_{\mathbf{k}}^{*}}\{b_{\mathbf{k}}^{*},H_{\mathrm{aux}}\}_{b}\right]\nonumber \\
 & = & -i\{O(b,b^{*}),H_{\mathrm{aux}}\}_{b}\;. \nonumber \\
\label{dobbdz}
\end{eqnarray}
Similarly one obtains for the second derivative%
\begin{eqnarray}
\left.\frac{d^{2}O(\tilde{b},\tilde{b}^{*})}{d^{2}z}\right|_{z=0} & = & \left.\frac{d}{dz}\left[\frac{dO(\tilde{b},\tilde{b}^{*})}{dz}\right]\right|_{z=0} \nonumber \\
&=&-i\left\{ \left.\frac{dO(\tilde{b},\tilde{b}^{*})}{dz}\right|_{z=0},H_{\mathrm{aux}}\right\} _{b}\nonumber \\
& = & (-i)^{2}\{\{O(b,b^{*}),H_{\mathrm{aux}}\}_{b},H_{\mathrm{aux}}\}_{b}\;, \nonumber \\
 \end{eqnarray}
such that by induction
\begin{eqnarray}
\left.\frac{d^{n}O(\tilde{b},\tilde{b}^{*})}{d^{n}z}\right|_{z=0} & = & \left.\frac{d}{dz}\left[\frac{d^{n-1}O(\tilde{b},\tilde{b}^{*})}{d^{n-1}z}\right]\right|_{z=0} \nonumber \\
&=&-i\left\{ \left.\frac{d^{n-1}O(\tilde{b},\tilde{b}^{*})}{d^{n-1}z}\right|_{z=0},H_{\mathrm{aux}}\right\} _{b}\nonumber \\
 & = & (-i)^{n}\{\{O(b,b^{*}),\underbrace{H_{\mathrm{aux}}\}_{b},\cdots,H_{\mathrm{aux}}}_{n\times H_{\mathrm{aux}}}\}_{b}\,, \nonumber \\
\label{eq:proof-h-new}
\end{eqnarray}
where we have used the shorthand notations $O(\tilde{b},\tilde{b}^{*})
\equiv O(\tb_{\mathbf{p}}(z),\tb_{\mathbf{p}}^{*}(z))$ 
and $O(b,b^*)\equiv O(\tilde{b}\rightarrow b,\tilde{b}^{*}\rightarrow b^*)$ when setting $z=0$ 
in $O(\tb_{\mathbf{p}}(z),\tb_{\mathbf{p}}^{*}(z))$. We have also implied in Eq. (\ref{eq:proof-h-new}) 
that $H_{\mathrm{aux}}\equiv H_{\mathrm{aux}}(b,b^*)=\left. H_{\mathrm{aux}}(\tilde{b},\tilde{b}^*)\right|_{z=0}$. 
We can then perform a Taylor expansion of $O(\tb_{\mathbf{p}}(z),\tb_{\mathbf{p}}^{*}(z))$ 
with respect to $z$ at $z=0$ whose coefficients are given by Eq. (\ref{eq:proof-h-new}). 
Setting $O=H$ we obtain Eq. (\ref{eq:h-new}).

\section{Derivation of coefficients in $H_{\mathrm{aux}}$}
\label{sec:coef}
We will provide here the detailed derivation
of the coefficients in Eq. (\ref{eq:solution-B}).
For convenience, we can write the interaction part of the Hamiltonian (\ref{eq:ham-int})
in the compact form (replacing all $a_{\mathbf{k}}$ with $b_{\mathbf{k}}$),
\begin{eqnarray}
&& H_{\mathrm{int}} (b,b^*)  =  \frac{\lambda}{24}\int \frac{[d^{3}\mathbf{k}_{1}][d^{3}\mathbf{k}_{2}][d^{3}\mathbf{k}_{3}][d^{3}\mathbf{k}_{4}]}{\sqrt{16E_{k1}E_{k2}E_{k3}E_{k4}}} (2\pi)^{3} \\
 &  & \times\sum_{\epsilon_{1},\epsilon_{2},\epsilon_{3},\epsilon_{4}=\pm1}b_{\mathbf{k}1}^{\epsilon_{1}}b_{\mathbf{k}2}^{\epsilon_{2}}b_{\mathbf{k}3}^{\epsilon_{3}}b_{\mathbf{k}4}^{\epsilon_{4}}\delta(\epsilon_{1}\mathbf{k}_{1}+\epsilon_{2}\mathbf{k}_{2}+\epsilon_{3}\mathbf{k}_{3}+\epsilon_{4}\mathbf{k}_{4}), \nonumber  \label{eq:h-int}
\end{eqnarray}
where we have used shorthand notations $b_{\mathbf{k}i}^{-}\equiv b_{\mathbf{k}i}$ and 
$b_{\mathbf{k}i}^{+}\equiv b_{\mathbf{k}i}^{*}$ with $i=1,2,3,4$.  
We can also write $H_{\mathrm{aux}}$ in Eq. (\ref{eq:h-aux}), using the same notation, in the compact form
\begin{eqnarray}
&& H_{\mathrm{aux}}  =  \frac{1}{24}\int[d^{3}\mathbf{k}_{1}][d^{3}\mathbf{k}_{2}][d^{3}\mathbf{k}_{3}][d^{3}\mathbf{k}_{4}]\nonumber \\
 &  & \times\sum_{\epsilon_{1},\epsilon_{2},\epsilon_{3},\epsilon_{4}=\pm1}^{\epsilon_{1}+\epsilon_{2}+\epsilon_{3}+\epsilon_{4} \ne 0}B^{\epsilon_{1}\epsilon_{2}\epsilon_{3}\epsilon_{4}}(\mathbf{k}_{1},\mathbf{k}_{2},\mathbf{k}_{3},\mathbf{k}_{4})b_{\mathbf{k}1}^{\epsilon_{1}}b_{\mathbf{k}2}^{\epsilon_{2}}b_{\mathbf{k}3}^{\epsilon_{3}}b_{\mathbf{k}4}^{\epsilon_{4}}\nonumber \\
 &  & \times\delta(\epsilon_{1}\mathbf{k}_{1}+\epsilon_{2}\mathbf{k}_{2}+\epsilon_{3}\mathbf{k}_{3}+\epsilon_{4}\mathbf{k}_{4}) \,,
\end{eqnarray}
where we have assumed $B^{----}(\mathbf{k}_{1},\mathbf{k}_{2},\mathbf{k}_{3},\mathbf{k}_{4})=B_{1}(\mathbf{k}_{1},\mathbf{k}_{2},\mathbf{k}_{3},\mathbf{k}_{4})$,
$B^{---+}(\mathbf{k}_{1},\mathbf{k}_{2},\mathbf{k}_{3},\mathbf{k}_{4})=B_{2}(\mathbf{k}_{1},\mathbf{k}_{2},\mathbf{k}_{3};\mathbf{k}_{4})$,
$B^{--++}=0$, $(B^{\epsilon_{1}\epsilon_{2}\epsilon_{3}\epsilon_{4}})^{*}=B^{-\epsilon_{1},-\epsilon_{2},-\epsilon_{3},-\epsilon_{4}}$
and $B^{\epsilon_{1}\epsilon_{2}\epsilon_{3}\epsilon_{4}}(\mathbf{k}_{1},\mathbf{k}_{2},\mathbf{k}_{3},\mathbf{k}_{4})$
is symmetric with respect to the exchange of any two labels, $\epsilon_{i}\leftrightarrow\epsilon_{j}$
and $\mathbf{k}_{i}\leftrightarrow\mathbf{k}_{j}$ with $i\neq j$. As we will leave the resonant term in Eq. (\ref{eq:h-int}) intact, 
there will be no resonant term in $H_{\mathrm{aux}}$, which requires $\epsilon_{1}+\epsilon_{2}+\epsilon_{3}+\epsilon_{4} \ne 0$. 

With the compact and symmetric form of $H_{\mathrm{int}}$ and $H_{\mathrm{aux}}$, it is easy to check 
\begin{eqnarray}
&& \{H_{0}(b,b^*),H_{\mathrm{aux}}\}_{b}  =  \frac{1}{24}\int[d^{3}\mathbf{k}_{1}][d^{3}\mathbf{k}_{2}][d^{3}\mathbf{k}_{3}][d^{3}\mathbf{k}_{4}]\nonumber \\
 &  & \times\sum_{\epsilon_{1},\epsilon_{2},\epsilon_{3},\epsilon_{4}=\pm}B^{\epsilon_{1},\epsilon_{2},\epsilon_{3},\epsilon_{4}}(\mathbf{k}_{1},\mathbf{k}_{2},\mathbf{k}_{3},\mathbf{k}_{4}) b_{\mathbf{k}1}^{\epsilon_{1}}b_{\mathbf{k}2}^{\epsilon_{2}}b_{\mathbf{k}3}^{\epsilon_{3}}b_{\mathbf{k}4}^{\epsilon_{4}}\nonumber \\
 &  & \times \quad ~\delta(\epsilon_{1}\mathbf{k}_{1}+\epsilon_{2}\mathbf{k}_{2}+\epsilon_{3}\mathbf{k}_{3}+\epsilon_{4}\mathbf{k}_{4})\nonumber \\
 &  & \times  \quad \left(\epsilon_{1}E_{k1}+\epsilon_{2}E_{k2}+\epsilon_{3}E_{k3}+\epsilon_{4}E_{k4}\right)\,,
\label{H0Haux4}
\end{eqnarray}
where we have used
\begin{eqnarray}
&&\left\{ b_{\mathbf{k}}b_{\mathbf{k}}^{*},b_{\mathbf{k}1}^{\epsilon_{1}}b_{\mathbf{k}2}^{\epsilon_{2}}b_{\mathbf{k}3}^{\epsilon_{3}}b_{\mathbf{k}4}^{\epsilon_{4}}\right\} _{b} = (2\pi)^{3}b_{\mathbf{k}1}^{\epsilon_{1}}b_{\mathbf{k}2}^{\epsilon_{2}}b_{\mathbf{k}3}^{\epsilon_{3}}b_{\mathbf{k}4}^{\epsilon_{4}} ~ \left[\epsilon_{1}\delta(\mathbf{k}_{1}-\mathbf{k}) \right. \nonumber \\
&&\left. \qquad +\epsilon_{2}\delta(\mathbf{k}_{2}-\mathbf{k})+\epsilon_{3}\delta(\mathbf{k}_{3}-\mathbf{k})+\epsilon_{4}\delta(\mathbf{k}_{4}-\mathbf{k})\right]. 
\label{comm-b2-b4}
\end{eqnarray}
One can use the Poisson bracket rules in Eq.~(\ref{eq:rule-poisson}) to establish the above formula.

To remove the non-resonant term in $H_{\mathrm{int}}$, it is natural to require 
\begin{equation}
H_{\mathrm{int}}^{\mathrm{non-res}}(b,b^*)+(-iz)\{H_{0}(b,b^*),H_{\mathrm{aux}}\}_{b} =0\, .
\end{equation}
The solution to the above equation is
\begin{eqnarray}
&& B^{\epsilon_1, \epsilon_2, \epsilon_3, \epsilon_4}(\mathbf{k_1},\mathbf{k_2},\mathbf{k_3},\mathbf{k_4})  = \nonumber\\
&& \qquad \frac{\lambda}{iz} \frac{(2\pi)^{3}}{\sqrt{16E_{k1}E_{k2}E_{k3}E_{k4}}} \nonumber\\
&& \qquad \times \frac{1}{\epsilon_1 E_{k1} +\epsilon_2 E_{k2} +\epsilon_3 E_{k3} +\epsilon_4 E_{k4}}\,, \nonumber\\
\end{eqnarray}
which explicitly give   
\begin{eqnarray}
&& B_{1}(\mathbf{k}_{1},\mathbf{k}_{2},\mathbf{k}_{3},\mathbf{k}_{4})  = \nonumber\\
&&  \qquad i\frac{1}{z}\lambda\frac{(2\pi)^{3}}{\sqrt{16E_{k1}E_{k2}E_{k3}E_{k4}}(E_{k1}+E_{k2}+E_{k3}+E_{k4})} \,, \nonumber\\
 && B_{2}(\mathbf{k}_{1},\mathbf{k}_{2},\mathbf{k}_{3};\mathbf{k}_{4})  = \nonumber\\
 && \qquad i\frac{1}{z}\lambda\frac{(2\pi)^{3}}{\sqrt{16E_{k1}E_{k2}E_{k3}E_{k4}}(E_{k1}+E_{k2}+E_{k3}-E_{k4})} \,,  \nonumber\\ 
\end{eqnarray}
with $B_1=B^{----}, B_1^*=B^{++++}, B_2=B^{---+}, B_2^*=B^{+++-}$.

\section{Properties of Bose superfluids from Gross-Pitaevskii Hamiltonian}
\label{sec:dispersion}
For completeness, we will review in this appendix the quasi-particle
dispersion relation as well as other properties of Bose superfluids extracted from the GP   
Hamiltonian in Eq.~(\ref{eq:ham-gp}).

Since there is no number changing channel in the interaction, 
the total particle number $N=\int[d^{3}\mathbf{p}]b_{\mathbf{p}}b_{\mathbf{p}}^{*}$
is conserved. It can be decomposed into contributions of zero and
non-zero modes. The zero mode corresponds to Bose condensation. Accordingly
we can write $b_{\mathbf{p}}$ in the following form, 
\begin{equation}
b_{\mathbf{p}}=\tilde{b}_{0}(2\pi)^{3}\delta(\mathbf{p})+b_{\mathbf{p}}^{\prime},\label{eq-b0}
\end{equation}
where $\tilde{b}_{0}$ and $b_{\mathbf{p}}^{\prime}$ correspond to
the zero and non-zero mode respectively. Note that $\tilde{b}_{0}$
can be set to a real number, as $b$ and $b^{*}$ always appear in
pairs and the phase of $\tilde{b}_{0}$ will be canceled if it has
any. So we have 
\begin{equation}
N=N_{0}+\int[d^{3}\mathbf{p}]b_{\mathbf{p}}^{\prime}b_{\mathbf{p}}^{\prime*}
\end{equation}
where $N_{0}=\tilde{b}_{0}^{2}V$ is the particle number of the zero
mode and $V$ is the volume. Inserting Eq.(\ref{eq-b0}) into Eq.(\ref{eq:ham-gp})
and keeping quadratic terms only, the Hamiltonian up to quadratic
terms in $b_{\mathbf{p}}^{\prime}$ and $b_{\mathbf{p}}^{\prime*}$
takes the form, 
\begin{eqnarray}
H & \rightarrow & mN+\frac{\lambda N_{0}^{2}}{16m^{2}V}+\int[d^{3}\mathbf{p}]\frac{|\mathbf{p}|^{2}}{2m}b_{\mathbf{p}}^{\prime}b_{\mathbf{p}}^{\prime*}\nonumber \\
 &  & +\frac{\lambda N_{0}}{16m^{2}V}\int[d^{3}\mathbf{p}]\left\{ b_{\mathbf{p}}^{\prime}b_{-\mathbf{p}}^{\prime}+b_{\mathbf{p}}^{\prime*}b_{-\mathbf{p}}^{\prime*}+4b_{\mathbf{p}}^{\prime}b_{\mathbf{p}}^{\prime*}\right\} \nonumber \\
 & \rightarrow & mN+\frac{\lambda N^{2}}{16m^{2}V}+\int[d^{3}\mathbf{p}]\frac{|\mathbf{p}|^{2}}{2m}b_{\mathbf{p}}^{\prime}b_{\mathbf{p}}^{\prime*}\nonumber \\
 &  & +\frac{\lambda N_{0}}{16m^{2}V}\int[d^{3}\mathbf{p}]\left\{ b_{\mathbf{p}}^{\prime}b_{-\mathbf{p}}^{\prime}+b_{\mathbf{p}}^{\prime*}b_{-\mathbf{p}}^{\prime*}+2b_{\mathbf{p}}^{\prime}b_{\mathbf{p}}^{\prime*}\right\} , \nonumber \\
 \label{eq:ham-condensate}
\end{eqnarray}
where the first term $mN$ is static mass energy and will not be considered
in a nonrelativistic theory. 

The non-relativistic Hamiltonian of quadratic terms can be diagonalized
by Bogoliubov transformation, $b_{\mathbf{p}}^{\prime}=u_{\mathbf{p}}c_{\mathbf{p}}+v_{\mathbf{p}}c_{-\mathbf{p}}^{*}$,
with canonical conditions $|u_{\mathbf{p}}|^{2}-|v_{\mathbf{p}}|^{2}=1$
and $u_{\mathbf{p}}v_{\mathbf{-p}}=u_{\mathbf{-p}}v_{\mathbf{p}}$.
Due to the symmetry in our setting, we may assume that $u_{\mathbf{p}}=u_{\mathbf{-p}}$,
$v_{\mathbf{p}}=v_{\mathbf{-p}}$. Then the diagonalization can be
completed by solving $u_{\mathbf{p}}$ and $v_{\mathbf{p}}$ from
the complex equations, 
\begin{eqnarray}
\left(E_{p}^{\mathrm{nr}}+\frac{1}{2}\Delta\right)\left(|u_{\mathbf{p}}|^{2}+|v_{\mathbf{p}}|^{2}\right)+\frac{1}{2}\Delta u_{\mathbf{p}}v_{\mathbf{p}}+\frac{1}{2}\Delta u_{\mathbf{p}}^{*}v_{\mathbf{p}}^{*} & = & \omega_{\mathbf{p}},\label{eq:complex1}\nonumber \\
\\
\left(E_{p}^{\mathrm{nr}}+\frac{1}{2}\Delta\right)u_{\mathbf{p}}v_{\mathbf{p}}^{*}+\frac{1}{4}\Delta u_{\mathbf{p}}u_{\mathbf{p}}+\frac{1}{4}\Delta v_{\mathbf{p}}^{*}v_{\mathbf{p}}^{*} & = & 0, \nonumber \\
\label{eq:complex2}
\end{eqnarray}
where we have defined $E_{p}^{\mathrm{nr}}=|\mathbf{p}|^{2}/(2m)$,
$\Delta=\lambda N_{0}/(4m^{2}V)$. The solution can be parametrized
as $u_{\mathbf{p}}=1-iz$ and $v_{\mathbf{p}}=y-ix$ with $z^{2}=x^{2}+y^{2}$
to guarantee the canonical conditions. 
Solving Eq. (\ref{eq:complex2}) we obtain the the dispersion relation,
\begin{equation}
\omega_{p}=\sqrt{E_{p}^{\mathrm{nr}}(E_{p}^{\mathrm{nr}}+\Delta)}\,. 
\label{eq:dis-non-rel}
\end{equation}
We can choose the positive energy solutionand thereby reproduce 
the Landau quasi-particle spectrum in Bose superfluids. 

There is another easier derivation of the quasi-particle energy (\ref{eq:dis-non-rel})
by using the canonical coordinate and momentum similar to Eq. (\ref{eq:ak-ask}),
\begin{eqnarray}
 &  & \phi_{\mathbf{k}}=\frac{1}{\sqrt{2E_{k}^{\mathrm{nr}}}}(b_{\mathbf{k}}^{\prime}+b_{-\mathbf{k}}^{\prime*}),\;\;\pi_{\mathbf{k}}=-i\sqrt{\frac{E_{k}^{\mathrm{nr}}}{2}}(b_{\mathbf{k}}^{\prime}-b_{-\mathbf{k}}^{\prime*}),\nonumber \\
 &  & \phi_{-\mathbf{k}}=\phi_{\mathbf{k}}^{*},\;\;\pi_{-\mathbf{k}}=\pi_{\mathbf{k}}^{*}.
\end{eqnarray}
Then the non-relativistic part of the Hamiltonian in (\ref{eq:ham-condensate})
becomes 
\begin{eqnarray}
H_{\mathrm{nr}} & = & \int[d^{3}\mathbf{k}]\frac{1}{2}\left(E_{p}^{\mathrm{nr}}+\frac{1}{2}\Delta\right)\left[E_{p}^{\mathrm{nr}}|\phi_{\mathbf{p}}|^{2}+\frac{1}{E_{p}^{\mathrm{nr}}}|\pi_{\mathbf{p}}|^{2}\right]\nonumber \\
 &  & +\frac{1}{4}\Delta\int[d^{3}\mathbf{p}]\left[E_{p}^{\mathrm{nr}}|\phi_{\mathbf{p}}|^{2}-\frac{1}{E_{p}^{\mathrm{nr}}}|\pi_{\mathbf{p}}|^{2}\right]\nonumber \\
 & = & \int[d^{3}\mathbf{p}]\frac{1}{2}\left[|\pi_{\mathbf{p}}|^{2}+E_{p}^{\mathrm{nr}}(E_{p}^{\mathrm{nr}}+\Delta)|\phi_{\mathbf{p}}|^{2}\right],
\end{eqnarray}
which is similar to a system of Harmonic oscillators with the frequency
or energy given in (\ref{eq:dis-non-rel}) for each momentum mode. 

In the relativistic theory, the relativistic energy should be taken into account.
From the Hamiltonian (\ref{eq:ham-new}), a similar dispersion relation
to Eq. (\ref{eq:dis-non-rel}) can also be obtained 
\begin{equation}
\omega_{p}=\sqrt{E_{p}\left(E_{p}+\frac{\lambda N_{0}}{4mE_{p}V}\right)},\label{eq:dis-rel}
\end{equation}
where $E_{p}=\sqrt{|\mathbf{p}|^{2}+m^{2}}$.


From Eq. (\ref{eq:ham-gp}), we obtain the Hamiltonian density of the GP field  
\begin{eqnarray}
\mathcal{H} & = & m|\psi|^{2}-\frac{1}{2m}\psi^{*}\nabla^{2}\psi+\frac{\lambda}{16m^{2}}|\psi|^{4}.\label{eq:energy-density}
\end{eqnarray}
The Lagrangian density can be obtained as  
\begin{eqnarray}
\mathcal{L} & = & \frac{1}{2}i\left(\psi^{*}\dot{\psi}-\psi\dot{\psi}^{*}\right)-\mathcal{H}.\label{eq:lagrangian}
\end{eqnarray}
In deriving equations of motion, we treat $\psi$ and $\psi^{*}$ as independent variables. 
It is interesting to see that the Hamiltonian and Lagrangian densities
are invariant under the global U(1) transformation $\psi(t,\mathbf{x})\rightarrow e^{i\eta}\psi(t,\mathbf{x})$.
There is no such symmetry in the original scalar field theory. For
the infinitesimal transformation $\eta\rightarrow0$, the global U(1)
invariance of the Lagrangian density leads to particle number conservation,
\begin{equation}
\frac{\partial}{\partial t}\rho+\nabla\cdot\mathbf{j}=0,\label{eq:charge-cons}
\end{equation}
where the particle number density and the current are given by 
\begin{eqnarray}
\rho & = & |\psi|^{2},\nonumber \\
\mathbf{j} & = & \frac{i}{2m}\left(\psi\nabla\psi^{*}-\psi^{*}\nabla\psi\right).
\end{eqnarray}
We can parameterize $\psi$ as $\psi=\sqrt{\rho}e^{i\phi}$ with $\rho$
being the particle number density which is assumed to be independent
of position. Hence the current is associated with the superfluid velocity
$\mathbf{v}=\frac{1}{m}\nabla\phi$, 
\begin{eqnarray}
\mathbf{j} & = & \rho\mathbf{v}.
\end{eqnarray}
We see that the phase $\phi$ acts as a potential for the fluid velocity.
We take a loop integral for $\mathbf{v}$ 
which gives the Feynman-Onsager quantization condition 
\begin{eqnarray}
\oint_{C}d\mathbf{l}\cdot\mathbf{v} & = & \int_{A}d\mathbf{S}\cdot(\nabla\times\mathbf{v})=\frac{1}{m}2l\pi,\quad(l=\mathrm{integer}) \nonumber \\
\label{eq:vortex}
\end{eqnarray}
If $\mathbf{v}$ is well-defined everywhere in the area $A$ bounded
by the loop $C$, we have $l=0$ and $\nabla\times\mathbf{v}=0$ everywhere.
If $\mathbf{v}$ is singular at one point, we conclude that $l\neq0$,
and the solutions satisfying Eq. (\ref{eq:vortex}) are vortex 
solutions for $\mathbf{v}$. 

The conservation of energy and momentum is a result of invariance of the
Lagrangian density under translation of time and position, $t^{\prime}=t+\epsilon$,
$\mathbf{x}^{\prime}=\mathbf{x}+\mathbf{c}$. 
For an infinitesimal transformation $\epsilon\rightarrow0$ and $|\mathbf{c}|\rightarrow0$
the variations of the fields are  
\begin{eqnarray}
\delta\psi & = & \epsilon\frac{\partial\psi}{\partial t}+\mathbf{c}\cdot\nabla\psi , \nonumber \\
\delta\psi^{*} & = & \epsilon\frac{\partial\psi^{*}}{\partial t}+\mathbf{c}\cdot\nabla\psi^{*} .
\label{eq:field-change}
\end{eqnarray}
The variation of the Lagrangian density is 
\begin{eqnarray}
\delta\mathcal{L} & = & \epsilon\left[\frac{\partial}{\partial t}\mathcal{H}-\frac{1}{2m}\nabla\cdot(\dot{\psi}\nabla\psi^{*}+\dot{\psi}^{*}\nabla\psi)\right]+\epsilon\frac{\lambda}{16m^{2}}\frac{\partial}{\partial t}|\psi|^{4}\nonumber \\
 &  & -c_{j}\left[\frac{1}{2}i\frac{\partial}{\partial t}(\psi\nabla_{j}\psi^{*}-\psi^{*}\nabla_{j}\psi)\right.\nonumber \\
 &  & \left.+\frac{1}{2m}\nabla_{i}(\nabla_{i}\psi^{*}\nabla_{j}\psi+\nabla_{i}\psi\nabla_{j}\psi^{*})\right].
\label{eq:drv-en-cons}
\end{eqnarray}
where we have inserted $\delta\psi$ and $\delta\psi^{*}$ from Eq.
(\ref{eq:field-change}) and applied the GP equation (\ref{eq:eom}). 
Under the invariance of the Lagrangian density, 
the first square bracket gives energy conservation, while the last
one gives momentum conservation. We can rewrite the second term in
the first square bracket as 
\begin{equation}
\mathcal{P}^{\prime}  =  m\rho\mathbf{v}\left(1+\frac{\lambda}{8m^{3}}\rho\right)+\frac{1}{2}m\rho\mathbf{v}^{2}\mathbf{v}\,,
\end{equation}
where we have used the GP equation (\ref{eq:eom}) and assumed that $\rho$
is a constant independent of position. Note that the term 
$\lambda/(16m^{2})\partial _t|\psi|^{4}$ in Eq. (\ref{eq:drv-en-cons}) 
cancels the $\lambda/(8m^{2})\rho^{2}\mathbf{v}$ term in $\mathcal{P}^{\prime}$
with the continuity equation (\ref{eq:charge-cons}). We can therefore define
the momentum density as 
\begin{equation}
\mathcal{P}=m\left(\rho+\frac{1}{2}\rho\mathbf{v}^{2}\right)\mathbf{v}.
\end{equation}
Similarly, the energy density can be cast into the following form,
\begin{equation}
\mathcal{H}=m\rho+\frac{1}{2}m\rho\mathbf{v}^{2}+\frac{\lambda}{16m^{2}}\rho^{2}. 
\label{eq:h-super}
\end{equation}
Then energy conservation equation now reads as 
\begin{equation}
\frac{\partial}{\partial t}\mathcal{H}+\nabla\cdot\mathcal{P}=0. 
\label{eq:em-cons-fluid}
\end{equation}
The last square bracket in Eq. (\ref{eq:drv-en-cons}) can be rewritten as 
\begin{equation}
\frac{\partial(\rho v_{j})}{\partial t}+\rho\nabla_{i}(v_{i}v_{j}) =  0, 
\end{equation}
which is nothing but the Euler equation. 

It is interesting to note that the momentum density is just the energy
density (excluding the $\rho^{2}$ term) times the superfluid velocity. The
last term $\lambda/(16m^{2})\rho^{2}$ of $\mathcal{H}$ in
Eq. (\ref{eq:h-super}) is negligible compared to $m\rho$ when $|\psi|^{2}/m^{3}=\rho/m^{3}\ll1$.
Then we have $\mathcal{H}\approx m[\rho+(1/2)\rho\mathbf{v}^{2}]$
and $\mathcal{P}=\mathcal{H}\mathbf{v}$. Applying the particle number
conservation equation (\ref{eq:charge-cons}), the energy conservation
equation (\ref{eq:em-cons-fluid}) becomes  
\begin{eqnarray}
\frac{\partial}{\partial t}\mathcal{H}_{\mathrm{nr}}+\nabla\cdot\mathcal{P}_{\mathrm{nr}} & = & 0\label{eq:en-mom-density}
\end{eqnarray}
where $\mathcal{H}_{\mathrm{nr}}=(1/2) m\rho\mathbf{v}^{2}$
and $\mathcal{P}_{\mathrm{nr}}=\mathcal{H}_{\mathrm{nr}}\mathbf{v}$
are the non-relativistic energy and momentum respectively. In the non-relativistic
limit $v\ll1$, $\mathcal{P}$ is approximately $m\rho\mathbf{v}$.

Another interesting observation is that we can read out the energy per
particle from $\mathcal{H}$ in Eq. (\ref{eq:h-super}), 
\begin{eqnarray}
\varepsilon & = & \frac{\partial\mathcal{H}}{\partial\rho}=m+\frac{1}{2}m|\mathbf{v}|^{2}+\frac{\lambda}{8m^{2}}\rho.
\end{eqnarray}
The above can be verified to be the same as the relativitic $\omega_{p}$
in Eq. (\ref{eq:dis-rel}) in the non-relativistic limit, 
\begin{eqnarray}
\omega_{p} & \approx & E_{p}+\frac{\lambda\rho}{8m^{2}}\approx\varepsilon,
\end{eqnarray}
where we have taken the limit $E_{p}\gg \lambda/(8m^{2}) \rho$
and made replacements $N_{0}/V\rightarrow\rho$, $\mathbf{p}\rightarrow m\mathbf{v}$
and then approximated $E_{p}\approx m+(1/2) m|\mathbf{v}|^{2}$.  

We now consider nonrelativistic case in which the superfluid momentum
density is $\mathcal{P}=m\rho\mathbf{v}$. Suppose a quasi-particle
with momentum $\mathbf{p}$ and energy $\omega_{p}$ is produced in 
the system, the momentum density changes to 
\begin{eqnarray}
\mathcal{P}^{\prime} & = & \mathcal{P}-\frac{\mathbf{p}}{V}=m\rho\mathbf{v}^{\prime},
\end{eqnarray}
where $\mathbf{v}^{\prime}=\mathbf{v}-\mathbf{p}/(mN)$ is the
new fluid velocity with $N=\rho V$ being the particle number in the
superfluid. Note that $N$ is very large, so we have $|\mathbf{p}|/(mN)\ll|\mathbf{v}|$.
The energy is changed due to the new velocity $\mathbf{v}^{\prime}$.
Hence according to Eq. (\ref{eq:en-mom-density}), the modified energy is 
\begin{eqnarray}
\mathcal{H}^{\prime} & = & \mathcal{H}+\frac{|\mathbf{p}|^{2}}{2mNV}-\mathcal{P}\cdot\frac{\mathbf{p}}{mN}\approx\mathcal{H}-\frac{\mathbf{v}\cdot\mathbf{p}}{V},
\end{eqnarray}
where we have neglected the term $|\mathbf{p}|^{2}/(2mNV)$ 
since it is much smaller than the rest terms. For a very small velocity,
the presence of quasiparticles violates energy conservation since
$\mathcal{H}^{\prime}>\mathcal{H}-\omega_{p}/V$. If the magnitude
of velocity is larger than a critical value $v_{c}=\omega_{p}/|\mathbf{p}|$,
the production of quasi-particles is possible since $\mathcal{H}^{\prime}=\mathcal{H}-\omega_{p}/V$
can be satisfied by making $\omega_{p}=\mathbf{v}\cdot\mathbf{p}$.
The condition for superfluids is that there exists a non-vanishing
critical velocity, $v_{c}>0$, which is called the Landau critical
velocity for superfluids. One can check that this is satified by the
quasi-particle energy (\ref{eq:dis-non-rel}) with $v_{c}=\sqrt{\lambda N_{0}/(8m^{3}V)}$.


\end{document}